\documentclass[12pt,twoside]{apa}
\usepackage{a4,amssymb,graphicx,eurosym}
\usepackage{apacite}
\usepackage{color}
\usepackage[utf8]{inputenc}
\usepackage[american]{babel}
\usepackage{amsmath}
\usepackage{amsfonts}
\usepackage{amssymb}
\usepackage{csquotes}
\usepackage{gensymb}
\usepackage{fancyhdr}
\usepackage{array}
\usepackage{dcolumn}
\usepackage{booktabs}
\usepackage{lineno}
\usepackage[doublespacing]{setspace}
\newcommand{\RNum}[1]{\uppercase\expandafter{\romannumeral #1\relax}}
\makeatletter
\long\def\@makecaption#1#2{%
  \vskip\abovecaptionskip
  \sbox\@tempboxa{#1 #2}%
  \ifdim \wd\@tempboxa >\hsize
    #1 #2\par
  \else
    \global \@minipagefalse
    \hb@xt@\hsize{\hfil\box\@tempboxa\hfil}%
  \fi
  \vskip\belowcaptionskip}
\makeatother

\definecolor{orange}{rgb}{1,0.5,0}

\usepackage{siunitx}
\usepackage{dcolumn}
\usepackage{graphicx}

\newcolumntype{L}[1]{>{\raggedright\arraybackslash}p{#1}} 
\newcolumntype{.}{D{.}{.}{-1}}
\newcolumntype{R}[1]{>{\raggedleft\arraybackslash}p{#1}} 

\usepackage{makecell}
\begin{document}
\thispagestyle{empty}
\pagenumbering{roman}
\begin{center}
{\huge\bf
The temporal evolution of the central fixation bias
in scene viewing}\\[0.3ex] 
\vspace{10mm}
\large
Lars O. M. Rothkegel$^{1*}$, Hans A.~Trukenbrod$^{1}$, Heiko H. Sch\"utt$^{1,2}$, \\[1ex]
Felix A.~Wichmann$^{2-4}$, and Ralf Engbert$^{1}$
\\
\vspace{10mm}
\normalsize
$^1$University of Potsdam, Germany\\
$^2$Eberhard Karls University T\"ubingen, Germany\\
$^3$Bernstein Center for Computational Neuroscience T\"ubingen, Germany\\
$^4$Max Planck Institute for Intelligent Systems, T\"ubingen, Germany\\
\vspace{10mm}
\today
\end{center}

\vspace{10mm}
\noindent
\vspace{\fill}

\noindent
$^*$To whom correspondence should be addressed: \\
Lars Rothkegel \\
Department of Psychology \& Cognitive Sciene Program  \\ 
University of Potsdam\\ 
Am Neuen Palais 10\\ 
14469 Potsdam\\
Germany \\ 
E-mail: lars.rothkegel\symbol{64}uni-potsdam.de \\
Phone: +49 331 9772370, Fax: +49 331 9772794

\newpage

\newpage
\setcounter{page}{1}
\pagenumbering{arabic}
\pagestyle{myheadings}
\markboth{Rothkegel et al.}{Temporal evolution of the central fixation bias}
\doublespacing

\section*{Abstract}
When watching the image of a natural scene on a computer screen, observers initially move their eyes towards the center of the image --- a reliable experimental finding termed {\sl central fixation bias}. This systematic tendency in eye guidance likely masks attentional selection driven by image properties and top-down cognitive processes. Here we show that the central fixation bias can be reduced by delaying the initial saccade relative to image onset. In four scene-viewing experiments we manipulated observers' initial gaze position and delayed their first saccade by a specific time interval relative to the onset of an image. We analyzed the distance to image center over time and show that the central fixation bias of initial fixations was significantly reduced after delayed saccade onsets. We additionally show that selection of the initial saccade target strongly depended on the first saccade latency.  Processes influencing the time course of the central fixation bias were investigated by comparing simulations of several dynamic and statistical models. Model comparisons suggest that the central fixation bias is generated by a default activation as a response to the sudden image onset and that this default activation pattern decreases over time. Our results suggest that it may often be preferable to use a modified version of the scene viewing paradigm that decouples image onset from the start signal for scene exploration and explicitly controls the central fixation bias. In general, the initial fixation location and the latency of the first saccade need to be taken into consideration when investigating eye movements during scene viewing.

\section{Introduction}

How humans visually explore natural scenes depends on multiple factors. Eye movements are influenced by low level image properties \cite<e.g., chromaticity, orientation, luminance, and color contrast;>{itti1998model,torralba2003modeling,le2006coherent} as well as higher level cognitive processes like the observers' scene understanding \cite{loftus1978cognitive,henderson1999effects}, task \cite{yarbus1967eye,castelhano2008influence}, or probability of reward \cite{hayhoe2005eye,tatler2011eye}. Besides low-level image features and high-level cognition, {\sl systematic tendencies} have a strong impact on how humans look at pictures \cite{tatler2009prominence,le2015saccadic}. A dominant systematic tendency in natural scene viewing is the {\sl central fixation bias}  \cite<CFB; >{buswell1935people,tatler2007central,tseng2009quantifying}. Regardless of stimulus material \cite{tatler2007central,tseng2009quantifying}, head position \cite{vitu2004eye}, initial fixation position \cite{tatler2007central,bindemann2010face}, or image position \cite{bindemann2010scene}, the eyes tend to initially fixate close to the center of an image when presented to a human observer on a computer screen. After several explanations of the CFB had been ruled out, two hypotheses remained. 

%


First, the image center might be the best location to maximize information extraction from scenes  \cite{najemnik2005optimal,tatler2007central} -- at least for typical photographs found in image databases and on the internet \cite<c.f;>{wichmann2010animal}. Second, the center of an image provides a strategic advantage to start inspection of a suddenly appearing stimulus \cite{tatler2007central}. Since realworld visual input doesn't suddenly appear and peripheral information of an upcoming stimulus is usually available, the CFB might be a laboratory artifact to some degree. Also, natural visual stimuli do not have rigid boundaries like a computer screen. A reduction of the CFB in mobile eye tracking data \cite{marius2009gaze, ioannidou2016centrial} supports this idea. 


A previous study from our lab resulted in a strong reduction of the CFB on initial fixations compared to similar experiments. In this study we manipulated the initial fixation by requiring participants to maintain fixation on a starting position close to the border of the screen for 1~s \cite{rothkegel2016influence}. In addition, some images in this study had asymmetric conspicuity distributions, with interesting or {\sl salient} image parts on either side of the image, but less so in the center. Thus, the reduction of the CFB in our scene viewing experiment could have been generated by three aspects: extreme initial starting positions, delayed initial saccades, and the saliency bias of the images we used.

To investigate the principles underlying the reduced CFB, we designed and analyzed four experiments, in which observers started exploration from different positions within an image and were required to maintain fixation for various time intervals after image onset (pre-trial fixation time). Our study used the images investigated in the most frequently cited paper on the central fixation bias \cite{tatler2007central}, in order to exclude any influence of the images on the reduction of the CFB. 


We hypothesized that (i) a forced prolonged initial fixation decouples image onset from the signal to start exploration and  leads to a reduced CFB on the second fixation which in turn reduces the bias on subsequent fixations \cite<due to the short saccade amplitudes of humans during scene perception; >{tatler2008systematic} and that (ii) the magnitude of the reduction varies with the duration of the prolonged initial fixation.

Here we show that the CFB of early eye movements can be reduced by dissociating initial eye movements from a sudden image onset by 125~ms and more. Increasing the delay of the initial response by more than 125~ms produced only marginal differences. In addition, we show that the initial saccade latency predicts the strength of the CFB on a trial-by-trial basis. The pre-trial fixation time primarily assures that the initial fixation is long enough to avoid a strong orienting response to the center of an image.

Finally we compare our data to five different models of saccade generation in scene viewing. An extended version of a recently published model of saccade generation \cite{engbert2015spatial} with an additional initial central activation that decreases over time best explains the data in terms of maximum likelihood \cite{schutt2016likelihood} and qualitative progression of the CFB over time. In terms of maximum likelihood a model that imitates the visual attention span of humans combined with the empirical fixation map explains the data equally well but fails to reproduce the qualitative progression of the CFB over time.

\section{General Methods}

\subsection{Stimuli}
A set of 120 images was presented  on a 20-inch CRT monitor (Mitsubishi Diamond Pro 2070: frame rate 120~Hz, resolution 1280$\times$1024 pixels; Mitsubishi Electric Corporation, Tokyo, Japan) in Experiment 1, 2 and 4 and on a different 20-inch CRT monitor (Iiyama Vision Master Pro 514: frame rate 100~Hz, resolution 1280$\times$1024 pixels; Iiyama, Nagano, Japan) in Experiment 3. The images were the same as in Tatler's (2007) original study on the central fixation bias. Images were indoor scenes (40 images), outdoor scenes with man-made structures present (e.g., urban scenes; 40 images), and outdoor scenes with no man-made structures present (40 images). Images were taken using a Nikon D2 digital SLR using its highest resolution (4 megapixel). All pictures had a size of 1600$\times$1200 pixels. For the presentation during the experiment, images were converted to a size of 1200$\times$900 pixels and centered on a screen with gray borders extending 64 pixels to the top/bottom and 40 pixels to the left/right of the image. In Experiments 1, 2 and 4 the images covered 31.1\degree~of visual angle in the horizontal and 23.3\degree~in the vertical dimension. In Experiment 3 images covered a larger proportion of the visual field with 36.25\degree~of visual angle in the horizontal and 27.20\degree~in the vertical dimension due to a reduced viewing distance.

\subsection{Participants}
Participants were students of the University of Potsdam and of nearby high schools. Number of participants for each experiment is indicated below. They received credit points or a monetary compensation of 8~Euro for their participation in any of the four experiments. The average duration of one experimental session was 40-45 minutes. All participants had normal or corrected-to-normal vision. The work was carried out in accordance with the Declaration of Helsinki. Informed consent was obtained for experimentation by all participants. 

\subsection{General Procedure}
Participants were instructed to position their heads on a chin rest in front of a computer screen at a viewing distance of 70~cm (60~cm in Exp.~3). Eye movements were recorded binocularly (monocularly in Exp.~3) using an Eyelink 1000 video-based-eyetracker (SR-Research, Osgoode/ON, Canada) with a sampling rate of 500~Hz (1000~Hz in Exp.~3 and downsampled to 500~Hz for our analysis). Trials began with a black fixation cross presented on gray background. After successful fixation, an image was presented. After onset of the image, the fixation cross remained visible on top of the image for a variable duration. We refer to this duration as the pre-trial fixation time. Participants were instructed to keep their eyes on the fixation cross until it disappeared. If participants moved their eyes before the pre-trial fixation time elapsed, a mask of Gaussian white noise was displayed and the trial started anew with the initial fixation check. After successful initial fixation, participants were instructed to explore the scene freely for five seconds in all experiments. Experiments were run with the Matlab software \cite{MATLAB:2015} using the Psychophysics  \cite{brainard1997psychophysics,pelli1997videotoolbox,kleiner2007s} and Eyelink \cite{cornelissen2002eyelink} toolboxes.


\subsection{Data Analysis} 
\subsubsection{Data preprocessing and saccade detection}

For saccade detection we applied a velocity-based algorithm \cite{engbert2003microsaccades,engbert2006microsaccades}. Saccades had a minimum amplitude of $0.5\degree$ and exceeded an average velocity during a trial by 6 (median-based) standard deviations for at least 6 data samples (12~ms). The epoch between two subsequent saccades was defined as a fixation. 

\subsection{Distance to Center over Time}
We computed the mean distance of the eye position to the image center $DTC$ as a function of pre-trial fixation time ($T$). This was computed as follows
\begin{equation}
\label{Eq_DTC}
{DTC}_{T} = \frac{1}{m\cdot n}\sum_{j=1}^n \sum_{k=1}^m {||x_{jk}(t)-x_{center}||} \;,
\end{equation}
where $x_{jk}(t)$ indicates gaze position of participant $j$ on image $k$ at time $t$ and $x_{center}$ indicates the image center. As a continuous-time measure, we computed the DTC of each sample of the eye-position time series. In this representation, a larger DTC indicates a less pronounced CFB and vice versa. For all experiments we visualized the mean ${DTC}(t) $ to the image center for the entire 5~s observation window for each pre-trial fixation time. The observation window started at $t=0$ with the disappearance of the fixation marker. All figures were created with the ggplot2 package \cite{ggplot} of the R-Language of Statistical Computing.


\subsubsection{Influence of the initial fixation on the second fixation}

The pre-trial fixation time influenced the DTC on early fixations. To further investigate this influence, we plot the DTC of the second fixation as a function of overall saccade latency from image onset. We computed linear mixed models \cite{bates2013lme4} with initial saccade latency and pre-trial fixation time as fixed effects, the DTC of the second fixation as the dependent variable and an intercept for subjects and images as random factors. To compute the models, we transformed DTC with the boxcox function of the R package MASS \cite{Venables:2002aa} to follow a normal distribution. We obtained significance levels with the lmerTest package \cite{kuznetsova2013lmertest}. Contrasts were defined as sum contrasts. This means that each pre-trial fixation time is compared to the overall mean of distance to center. To be able to compare the different factor levels to the overall mean, the highest pre-trial fixation time in each experiment was left out. In all experiments we excluded saccades with a latency smaller than or equal to 80~ms as anticipatory.

\subsubsection{Density Maps of eye positions over time}
To visualize the temporal evolution of eye positions in our experiments, we computed movies of 2D density maps for the different pre-trial fixation times and each eye position of the time-series recorded for each experiment. Based on a kernel density estimation via diffusion \cite{kde2d}, we estimated density maps for the first two seconds (after removal of the fixation cross) in each experiment. These movies are available as supplementary material.

\section{Experiment 1}
\subsection{Methods}
\subsubsection{Participants}
We recorded eye movements from 40 participants in Experiment 1 (34 female, 14--39 years old; 2 from a nearby high school)

\subsubsection{Procedure}
\begin{figure}
\unitlength1mm
\begin{picture}(150,73)
\put(30,-6){\includegraphics[width=12cm]{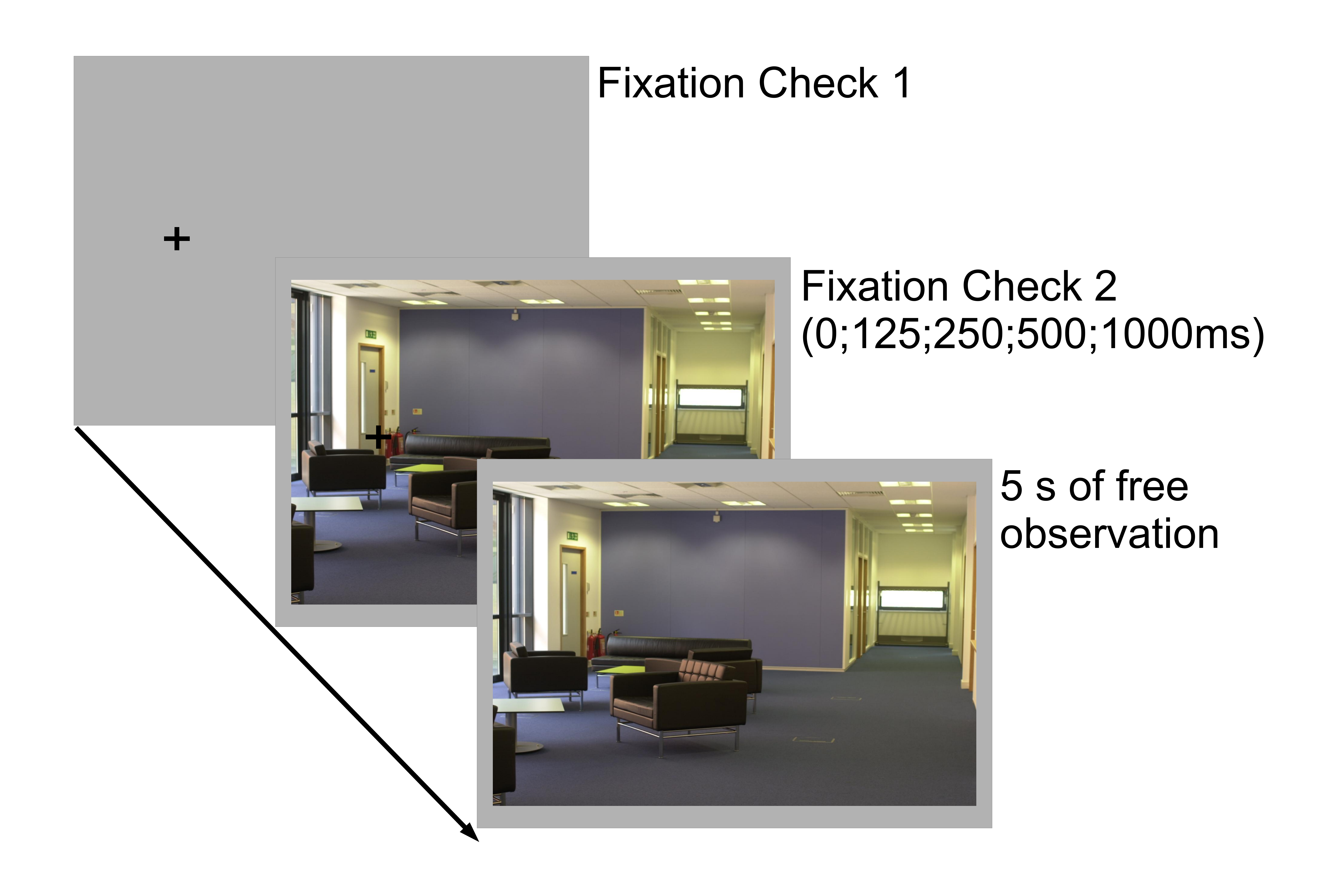}}
\end{picture}
\caption{\label{FigProcedure}
Schematic illustration of the experimental procedure of Experiment 1 with a starting position close to the left border of the screen. After a short fixation check of 200~ms (Fixation Check 1) the image is presented. A second fixation check between 0 and 1000~ms controls if participants move their eyes after image onset. After a succesful second fixation check, participants are allowed to freely move their eyes.}
\end{figure}

In Experiment 1 the fixation cross was presented at the horizontal meridian 5.6\degree~(256 Pixels) away from the left or right border of the monitor. This position was chosen to reproduce the findings of a strongly reduced central fixation bias observed in an earlier study \cite{rothkegel2016influence}, where participants experienced a pre-trial fixation time of 1~s. A proportion of 20\% of participants explored the image immediately after successful fixation without an additional pre-trial fixation time (0~ms). This corresponds to the standard scene viewing paradigm. For all other participants the fixation cross remained on top of the image for a duration of 125~ms, 250~ms, 500~ms, or 1000~ms. Pre-trial fixation time was used as a between-subject factor, i.e., each participant was tested with one of five pre-trial fixation times. Figure \ref{FigProcedure} illustrates a representative trial with the starting position on the left side of the screen. Fixation Check 2 was nonexistent for participants with a 0~ms pre-trial fixation time.
 
\subsection{Results}

\subsubsection{Distance to Center over Time}
\begin{figure}
\unitlength1mm
\begin{picture}(150,80)
\put(0,0){\includegraphics[width=7cm]{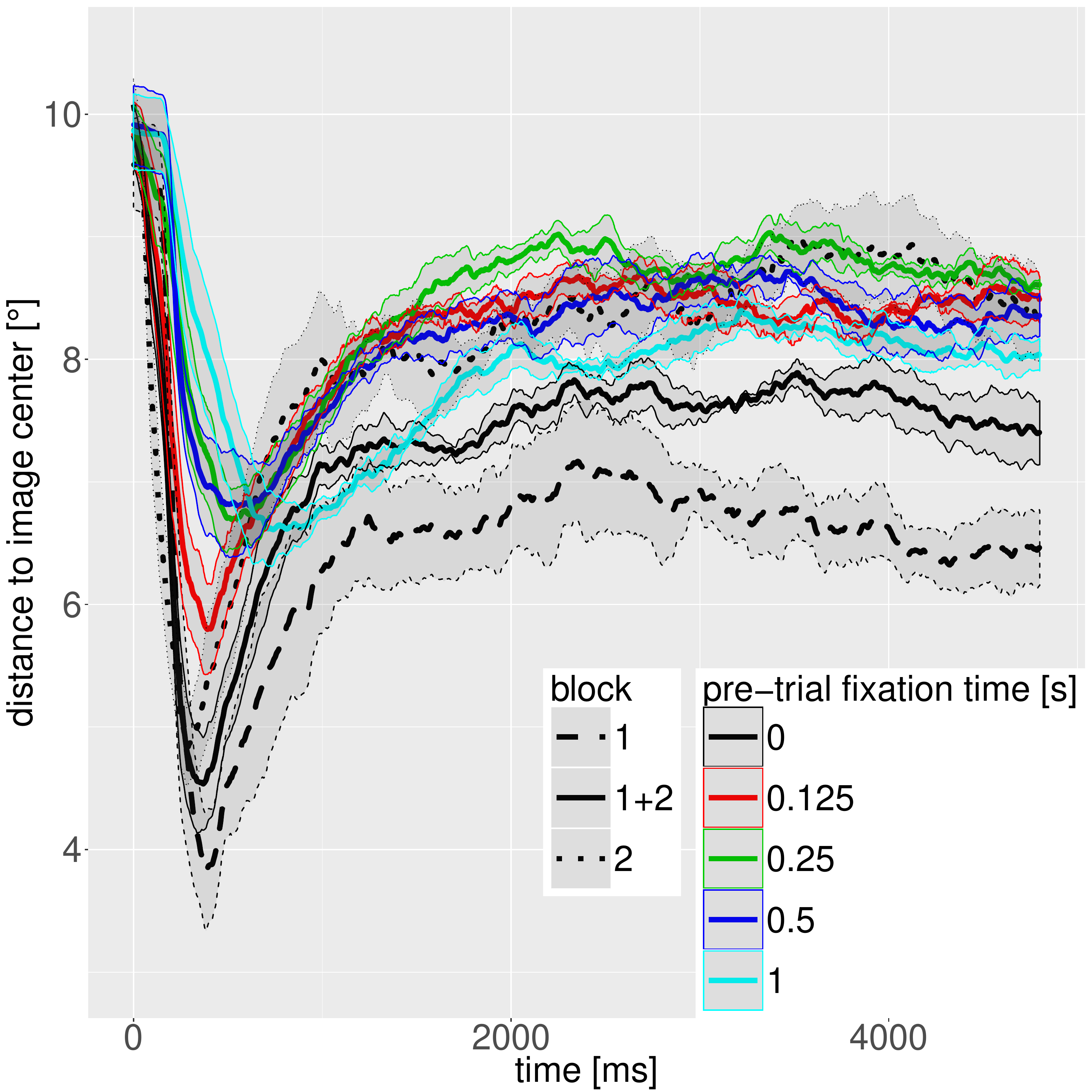}}
\put(80,0){\includegraphics[width=7cm]{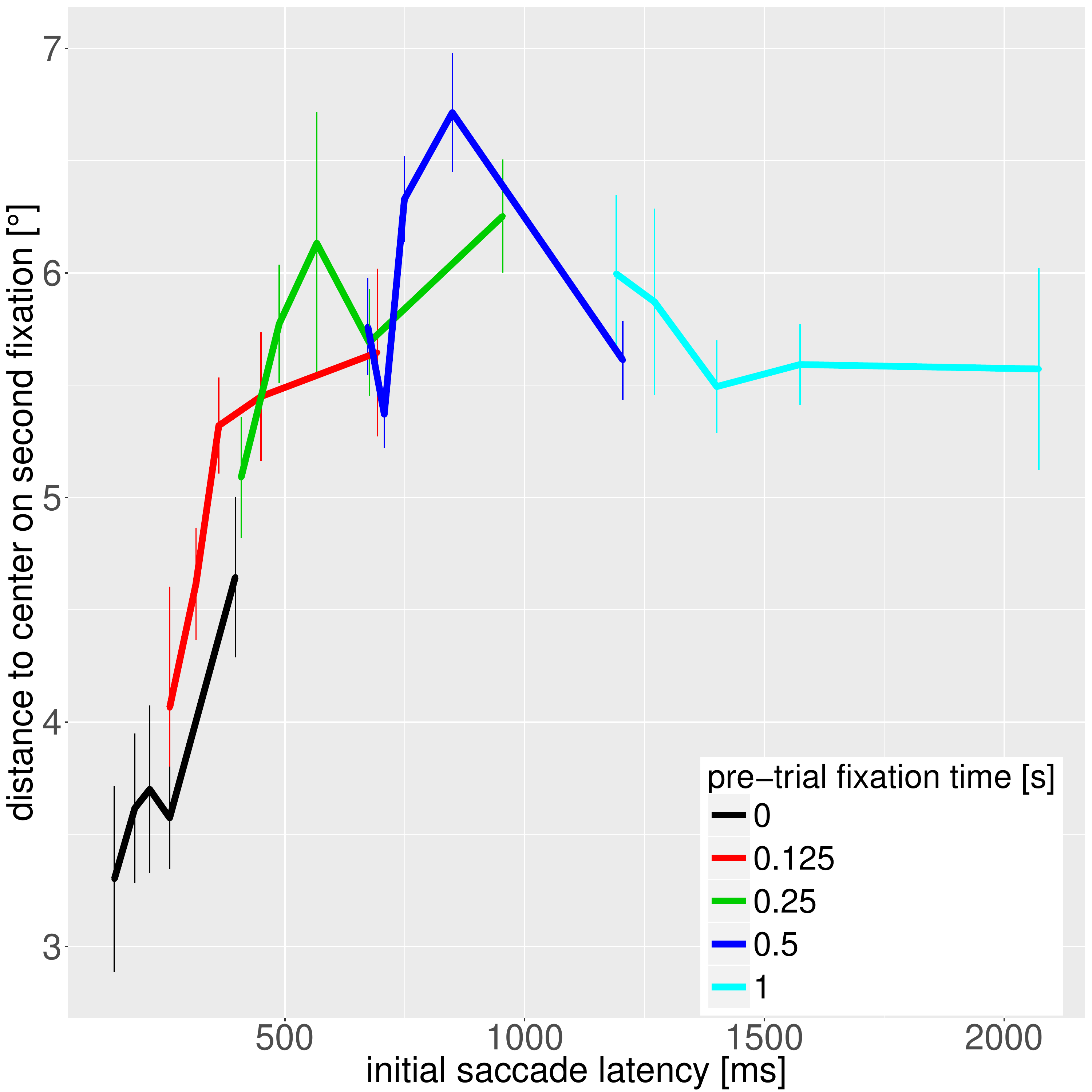}}
\put(0,75)a
\put(80,75)b
\end{picture}
\caption{\label{EXP1} Experiment 1. a) Mean distance to center over time (${DTC}(t)$) for the five different pre-trial fixation times with starting positions close to the border of the screen. Confidence intervals indicate standard errors as described by \protect \citeA{cousineau2005confidence}. Block 1 represents participants 1-20, Block 2 participants 21-40 who were originally tested as a follow up experiment to consolidate the results. b) Mean distance to center of the second fixation as a result of initial fixation duration. Error bars are 95\% confidence intervals of 1000 bootstrap samples as suggested by \protect \citeA{efron1994introduction}.}

\end{figure}

In Experiment 1, the ${DTC}$ initially decreased for all conditions (i.e., the CFB increased). There was a pronounced effect that mean fixation positions tended to be closer to the image center when participants were allowed to explore an image immediately after image onset, i.e., with a pre-trial fixation time of 0~ms (black curves in Fig.~\ref{EXP1}a). Surprisingly, for the first 4 participants (Block 1) of this group the effect was visible throughout the whole observation time of 5~s. A second group of participants in the 0~ms condition (Block 2) did not replicate the stronger CFB through the whole observation time. In addition, there was a gradual reduction of the CFB for pre-trial fixation times from 125~ms to 250~ms (red and green curve). ${DTC}$ for pre-trial fixation times of 250 and 500~ms hardly differed (green vs. blue curve). The minimum for the pre-trial fixation time of 1000~ms occured later in time because of disproportionately long saccade latencies of the first saccade after a forced fixation on the fixation cross of 1000~ms (cyan curve).

\subsubsection{Distance to Center on the second fixation}

Figure \ref{EXP1}b shows the influence of initial saccade latency on the mean DTC of the second fixation for the five pre-trial fixation times. Each bin represents a quintile of the distribution of saccade latencies in each condition. A clear relation between DTC of the second fixation and latency of the initial saccade is visible for the pre-trial fixation times of 0~ms and 125 ~ms. Overall, short saccade latencies led to a small average DTC (i.e., a strong initial CFB) whereas long latencies led to a larger average DTC (i.e., a less pronounced initial CFB).

Table \ref{LMMExp1} shows the output of the LMM for Experiment 1. The DTC for a pre-trial fixation time of 0~ms is significantly lower than the average DTC and for a pre-trial fixation time of 500~ms it is significantly higher. This means that a saccade immediately after a sudden image onset led to the strongest overall CFB in this experiment. The initial saccade latency is highly significant regardless of the pre-trial fixation time. The model also shows that an interaction between saccade latency and pre-trial fixation time exists. If the image appears directly after a succesfull fixation check (pre-trial fixation of 0~ms) the influence of saccade latency is significantly higher than on average (see row saccade latency $\times$ 0~ms). If pre-trial fixation time is as long as 500~ms, the influence of saccade latency is significantly weaker than on average (see line saccade latency $\times$ 500~ms). This interaction suggests that after a certain threshold time is reached, the influence of increasing saccade latency disappears.

\begin{table}
\caption{Output of LMM for Experiment 1}
\label{LMMExp1}
\small

\begin{tabular}{ @{} R{6.5 cm} . . .}
Fixed Effect   & \multicolumn{1}{c}{ Estimate }  &  \multicolumn{1}{c}{SE}  &\multicolumn{1}{c}{t} \\

\hline 
(Intercept) &  1.856  &  0.079 &    23.546^{***} \\
0~ms 				 &  -0.925 &  0.151 &  -6.106^{***} \\
125~ms   		&   -0.148 &  0.140 &   -1.053  \\
250~ms   		 &  0.185 &  0.140 & 1.320  \\
500~ms  				&  0.496 &  0.136 &  3.660^{***}  \\ 
saccade latency &  0.751   &  0.108  & 6.951^{***}  \\
saccade latency x 0~ms  & 1.685&   0.339 &  4.976^{***}  \\
saccade latency x 125~ms &  0.245 &    0.208 &  1.178    \\
saccade latency x 250~ms &  -0.210 &  0.175 &   -1.199   \\
saccade latency x 500~ms  & -0.886 &    0.155 &  -5.726^{***}   \\
\hline
\hline
Random effects variance: Subjects &   \multicolumn{1}{l}{  }  &  \multicolumn{1}{.}{0.1498} &\multicolumn{1}{l}{}  \\
Random effects variance: Images &   \multicolumn{1}{l}{  }  &  \multicolumn{1}{.}{0.1477}  &\multicolumn{1}{l}{}  \\
Log-Likelihood &   \multicolumn{1}{l}{  }  &  \multicolumn{1}{r}{-7135.53} &\multicolumn{1}{l}{}  \\
Deviance&     \multicolumn{1}{l}{  }  &  \multicolumn{1}{r}{14271.07}&\multicolumn{1}{l}{}   \\
AIC &     \multicolumn{1}{l}{  }  &  \multicolumn{1}{r}{14297.07}&\multicolumn{1}{l}{}   \\
BIC &     \multicolumn{1}{l}{  }  &  \multicolumn{1}{r}{14380.64} &\multicolumn{1}{l}{}  \\
N &     \multicolumn{1}{l}{  }  &  \multicolumn{1}{r}{4575} &\multicolumn{1}{l}{}\\[.2cm]
\multicolumn{4}{r}{ $~^{*} p<.05,~^{**} p<.01,~^{***} p<.001$} 
\end{tabular}

\end{table}


\subsection{Discussion}
Experiment 1 led to a reduction of the CFB on the initial saccade target for all pre-trial fixation times of 125~ms and more during scene perception from extreme starting positions (Fig.~\ref{EXP1}a). A pre-trial fixation time of 125~ms produced an intermediate CFB, whereas longer pre-trial fixation times produced asymptotic behavior. With a pre-trial fixation time of 0~s the DTC was smaller throughout almost the whole observation time of 5~s for the first group of participants. However, this effect was not replicated in a retest with 20 new participants. In this experiment the long influence of the pre-trial fixation time indeed vanished. The early effect of the CFB did not differ in the two groups of participants. The CFB of the second fixation did strongly depend on the latency of the initial saccade (Fig.~\ref{EXP1}b). Thus, the early differences between pre-trial fixation times in Figure \ref{EXP1}a are driven by differences in the distribution of intial saccade latencies. 

These results replicated our earlier findings of a reduced CFB during scene perception by introducing a non-zero pre-trial fixation time \cite{rothkegel2016influence}. A delay of 125~ms seemed sufficient to achieve a considerable reduction. In addition, our results suggested that the most important mediating factor of the CFB was the latency of the first saccadic response. Saccades with brief saccade latencies were on average directed more strongly towards the center than saccades with long saccade latencies.

\section{Experiment 2}
 
To assure that our results from Experiment 1 were not mainly induced by the extreme starting positions we conducted another experiment with starting positions closer to the image center.

\subsection{Methods}

\subsubsection{Participants}
We recorded eye movements from 20 participants for Experiment 2 (17 female; 14-28 years old; 1 from a nearby high school).

\subsubsection{Procedure}
Experiment 2 was similar to Experiment 1 except that the fixation cross was presented on a donut shaped ring with a distance of 2.6\degree~to 7.8\degree~(100 to 300 pixels) to the center. We used this donut shaped ring to obtain intermediate starting positions neither too close nor too far away from the center so that fixations could be directed both towards and away from the center. In addition, the donut shaped ring of starting positions made the initial starting position less predictable. This setup was thus slightly different to the experiment conducted by \citeA{tatler2007central} where the initial starting position was randomly chosen from a circle (fixed radius) around the image center. 

 
\subsection{Results}

\begin{figure}
\unitlength1mm
\begin{picture}(150,80)
\put(0,0){\includegraphics[width=7cm]{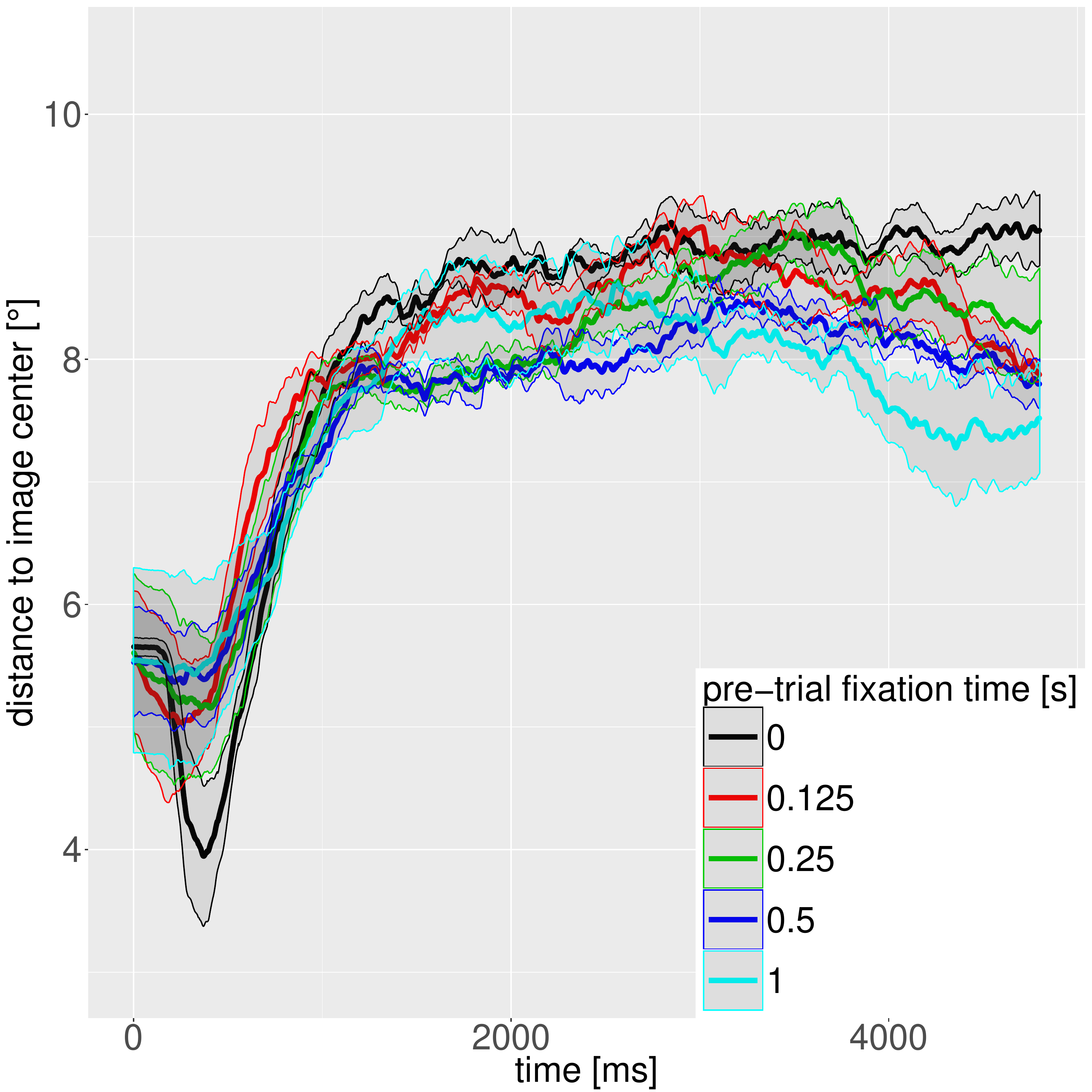}}
\put(80,0){\includegraphics[width=7cm]{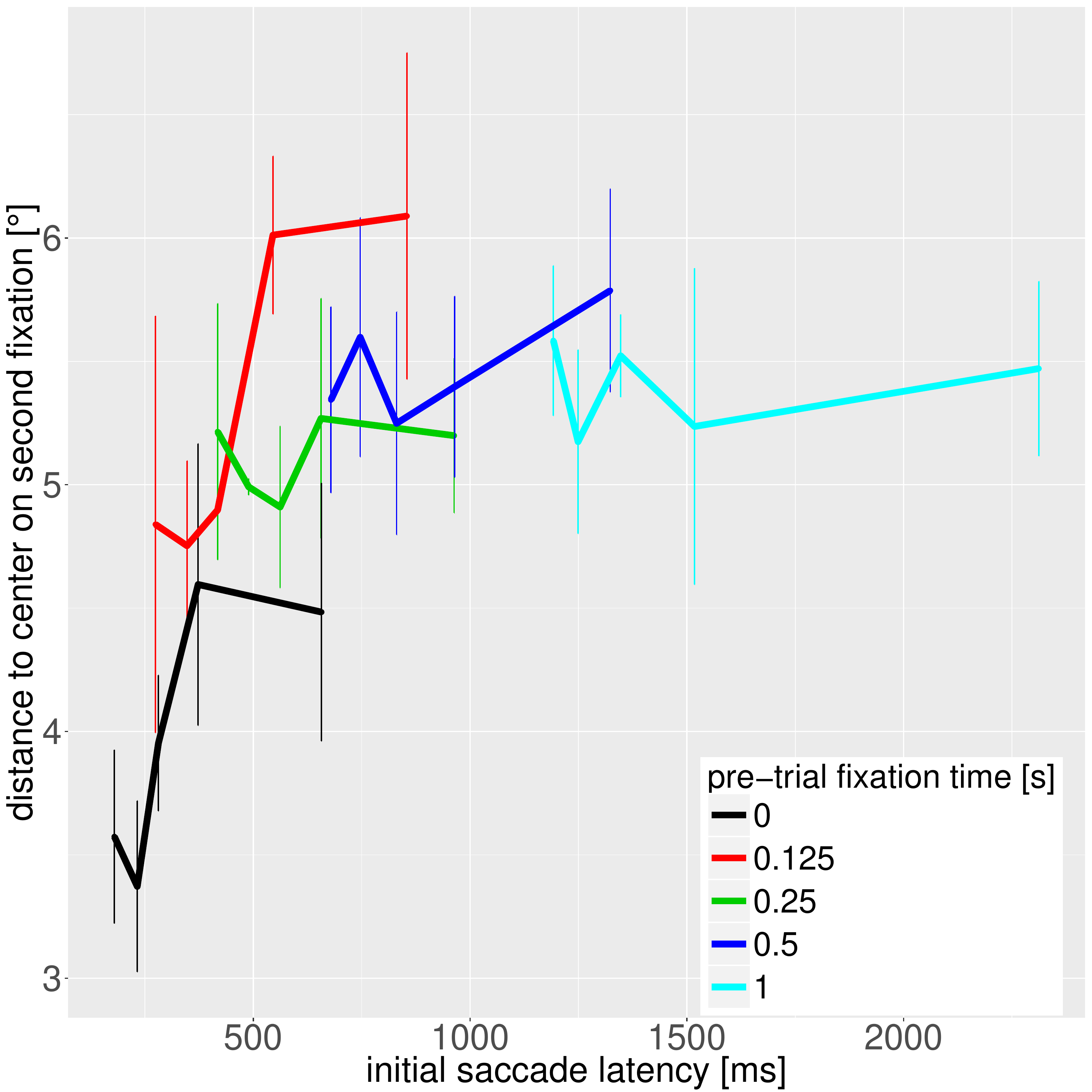}}
\put(0,75)a
\put(80,75)b
\end{picture}
\\
\caption{Experiment 2. a) Mean distance to center over time (${DTC}(t)$) for the five different pre-trial fixation times with starting positions on a donut shaped ring around the image center. Confidence intervals indicate standard errors as described by \protect \citeA{cousineau2005confidence}. b) Mean distance to center of the second fixation as a result of initial saccade latency and pre-trial fixation time. Bins represent quintiles of the saccade latency distribution. Errorbars are the standard error of the mean.} 
\label{EXP2}
\end{figure}

\subsubsection{Distance to Center over Time}
In Experiment 2, where the starting positions were located on a ring around the image center, the eyes moved initially even further towards the image center in the 0~ms pre-trial condition (black curve in Fig.~\ref{EXP2}a was the only curve with a pronounced negative slope in the beginning). A difference in ${DTC}$ was visible until about 600~ms after offset of the fixation marker. Later during the trial, the curves converged for all pre-trial conditions and reached a stable DTC for the rest of the trial. Qualitatively, we also observed a small initial difference in DTC between short pre-trial fixation times of 125~ms and 250~ms and pre-trial fixation times of 500~ms and 1000~ms.

\subsubsection{Distance to Center of the second fixation}
 As in Experiment 1, we found a strong influence of the latency of the first saccade on the DTC of the second fixation for small pre-trial fixation times (Fig.~\ref{EXP2}b). 
The results of the linear mixed model in Experiment 2 (Tab. \ref{LMMExp2}) were similar to Experiment 1. The most important results are the significantly lower DTC of the 0~ms pre-trial fixation time compared to the average and the significant increase in DTC for higher saccade latencies. As in Experiment 1 an interaction between saccade latency and pre-trial fixation time is visible. This is especially true for the 0~ms condition, where the influence of saccade latency significantly increases compared to the average influence. In Experiment 2 the only significant decrease in saccade latency influence is visible for a pre-trial fixation time of 250~ms. The overall direction of the influence (increasing influence of saccade latency for pre-trial fixation times of 0 \& 125~ms vs. decreasing influence for pre-trial fixation times of 250 \& 500~ms) remains as in Experiment 1. 

\begin{table}
\caption{Output of LMM for Experiment 2}
\label{LMMExp2}

\small

\begin{tabular}{ @{} R{6.5 cm} . . .}
Fixed Effect   & \multicolumn{1}{c}{ Estimate }  &  \multicolumn{1}{c}{SE}  &\multicolumn{1}{c}{t} \\

\hline 
(Intercept) &   2.167 & 0.094  & 22.965^{***} \\
0~ms 				 &  -0.899 &   0.186 &   -4.829^{***} \\
125~ms   		&   -0.010 &   0.177 &   -0.054  \\
250~ms   		 &  0.228 &   0.174 &    1.308  \\
500~ms  				&   0.255 &  0.177 & 1.439  \\ 
saccade latency &  0.468 &   0.110 &   4.258^{***}  \\
saccade latency x 0~ms  & 1.126 & 0.291 &  3.870^{***} \\
saccade latency x 125~ms &   0.327 &  0.229 &  1.430   \\
saccade latency x 250~ms &  -0.541 &  0.204 & -2.656^{**}    \\
saccade latency x 500~ms  & -0.235  & 0.208 &  -1.132 \\
\hline
\hline
Random effects variance: Subjects &   \multicolumn{1}{l}{  }  &  \multicolumn{1}{.}{0.1112}  &\multicolumn{1}{l}{}  \\
Random effects variance: Images &   \multicolumn{1}{l}{  }  &  \multicolumn{1}{.}{0.1333}  &\multicolumn{1}{l}{}  \\
Log-Likelihood &   \multicolumn{1}{l}{  }  &  \multicolumn{1}{.}{-3599.60}  &\multicolumn{1}{l}{}  \\
Deviance&     \multicolumn{1}{l}{  }  &  \multicolumn{1}{.}{7199.20}  &\multicolumn{1}{l}{}  \\
AIC &     \multicolumn{1}{l}{  }  &  \multicolumn{1}{.}{7225.20}  &\multicolumn{1}{l}{}  \\
BIC &     \multicolumn{1}{l}{  }  &  \multicolumn{1}{.}{7299.56}  &\multicolumn{1}{l}{} \\
N &     \multicolumn{1}{l}{  }  &  \multicolumn{1}{.}{2253}  &\multicolumn{1}{l}{} \\[.2cm]

\multicolumn{4}{r}{ $~^{*} p<.05,~^{**} p<.01,~^{***} p<.001$}

\end{tabular}
\end{table}


\subsection{Discussion}

If the starting position was close to the image center all pre-trial fixation times of 125~ms or longer (Fig.~\ref{EXP2}a) led to a reduction of the CFB on early fixations. After around 600~ms this influence disappeared. Furthermore, a clear relation between latency of the first saccade and the CFB of the second fixation was visible (Fig.~\ref{EXP2}b). Thus, the results replicated our observations from Experiment 1 and demonstrated that a reduced CFB was not exclusively generated by the extreme starting positions used in Experiment~1. 

\section{Experiment 3}
The results from Experiment 1 and 2 showed that a pre-trial fixation time of 125~ms was enough to reduce the central fixation bias on early fixations. The difference of the CFB between pre-trial fixation times larger than 125~ms was relatively small. To investigate the minimum pre-trial fixation time for a substantial CFB reduction, we conducted a third experiment with pre-trial fixation times ranging from 0 to 125~ms in six equidistant steps. We changed the between-subject design of pre-trial fixation time to a within-subject design to reduce the influence of individual participants (cf., Exp.~1). Hence, every participant was tested with all pre-trial fixation times. Since effects were maximal in the first experiment we used the same extreme starting positions as in Experiment 1.  


\subsection{Methods}

\subsubsection{Participants}
We recorded eye movements from 24 participants for Experiment 3 (20 female; 20--29 years old)
 
\subsubsection{Procedure}
In Experiment 3, participants experienced pre-trial fixation times between 0 and 125~ms in steps of 25~ms  (0, 25, 50, 75, 100, 125~ms). Each of the six pre-trial fixation times was presented in a block of 20 images, pseudo-randomized across participants. In total participants viewed 120 images. Note, the experiment was tested with a different setup (monitor, eye-tracker, etc.; see general methods section for details). Thus, the absolute value of DTC is not directly comparable between Experiment 3 and the remaining experiments. 

\subsection{Results}
\subsubsection{Distance to Center over Time}

\begin{figure}
\unitlength1mm
\begin{picture}(150,80)
\put(0,0){\includegraphics[width=7cm]{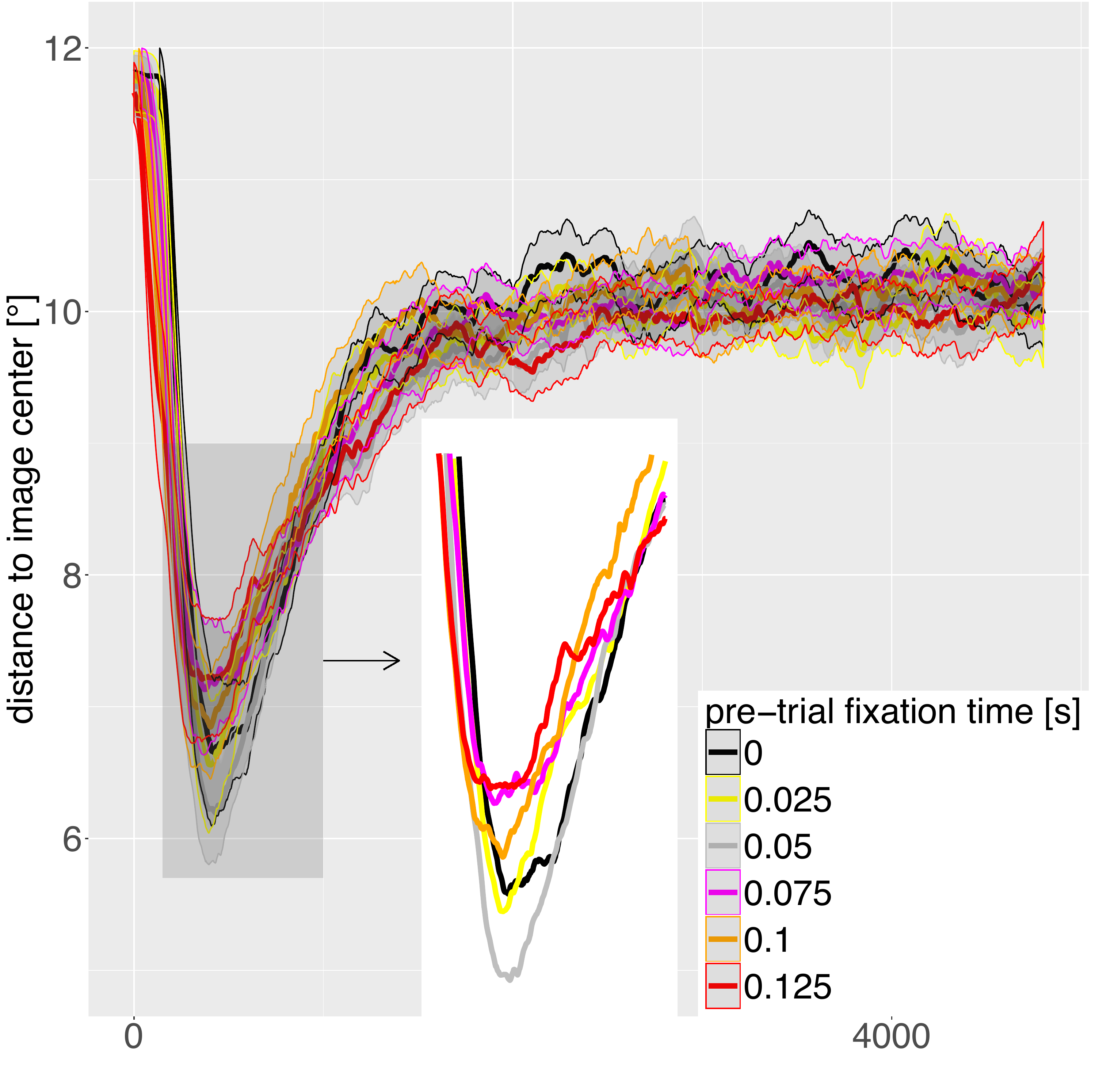}}
\put(80,0){\includegraphics[width=7cm]{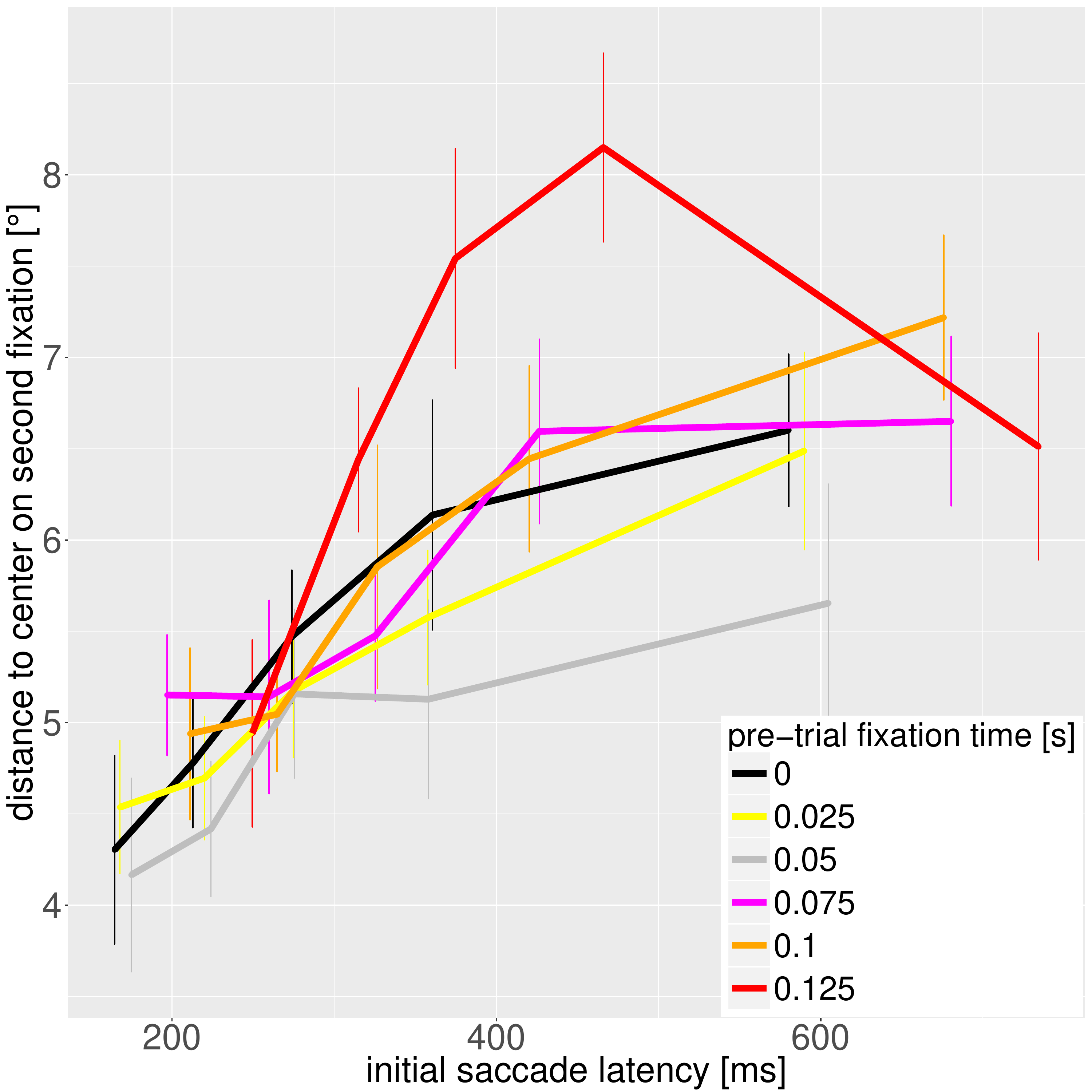}}
\put(0,75)a
\put(80,75)b
\end{picture}
\caption{Experiment 3. a) Mean distance to center over time (${DTC}(t)$) for the six different pre-trial fixation times with starting positions close to the left and right border. Confidence intervals indicate standard errors as described by \protect \citeA{cousineau2005confidence}. b) Mean distance to center of the second fixation as a result of initial saccade latency and pre-trial fixation time. Bins represent quintiles of the saccade latency distribution. Errorbars are the standard error of the mean.}
\label{EXP3}
\end{figure}

As in Experiment 1 and 2 the eyes initially moved towards the center for all pre-trial fixation times (Fig.~\ref{EXP3}a). The difference between pre-trial conditions was not as clearly visible as in previous experiments. Even the difference between the 0 and the 125~ms condition was relatively small. The smaller difference was probably due to the blocked design where pre-trial fixation times changed after 20 trials during the experiment for each participant. Nonetheless, curves with a pre-trial fixation time smaller than or equal to 50~ms had smaller minima than the ones with pre-trial fixation times larger than 50~ms (see inset in Fig.~\ref{EXP3}a).

\subsubsection{Distance to Center of the second fixation}

The influence of the first saccade latency on the distance to center of the second fixation is clearly visible in Figure \ref{EXP3}b. The influence seemed even clearer than in previous experiments. However, the range of the distance to center values was larger in this experiment as a result of the increased magnitude of the image in visual degree. Saccade latencies were more homogeneous in Experiment 3. The difference of mean saccade latencies (pre-trial fixation time + saccade latency after removal of the fixation marker) between the 0 and 125~ms condition was much smaller (57~ms) than in Experiments 1 (154~ms) and 2 (138~ms).

A linear mixed model for Experiment 3 showed that DTC of the second fixation did not show an independent influence of pre-trial fixation time (Tab. \ref{LMMExp3}). However, we replicated a significant influence of the first saccade latency on DTC of the second fixation. Shorter saccade latencies led to fixations closer to the center of an image. An interaction between pre-trial fixation time and saccade latency was not observed.

The saccade latencies of Experiment 3 were rather similar between the different pre-trial fixation times. There was a clear difference between the three lowest pre-trial fixation times (mean saccade-latencies of 315, 320 \&~321 ms) compared to the three longer pre-trial fixation times (mean saccade-latencies of 365, 352 \&~371 ms). Thus somewhere around 75~ms seems to be the lowest pre-trial fixation time to produce an influence in viewing behaviour.

\begin{table}
\caption{Output of LMM for Experiment 3}
\label{LMMExp3}
\small
\begin{tabular}{ @{} R{6.5 cm} . . .}

Fixed Effect   & \multicolumn{1}{c}{ Estimate }  &  \multicolumn{1}{c}{SE}  &\multicolumn{1}{c}{t}  \\

\hline 
(Intercept) &   2.187 &     0.110 &  19.832^{***}  \\
0~ms 				 &  0.056 &    0.076 & 0.731  \\
25~ms   		&     -0.096 &    0.086 & -1.118 \\
50~ms   		 &   -0.069 &   0.072 &   -0.949   \\
75~ms  				&    0.055 &     0.081 & 0.673 \\ 
100~ms  				&     -0.015 &    0.078 &  -0.190   \\ 
saccade latency &    1.056 &    0.110 & 9.564^{***}  \\
saccade latency x 0~ms  & -0.334 &  0.201 &  -1.662 \\
saccade latency x 25~ms &  0.161 &    0.249  & 0.645   \\
saccade latency x 50~ms &  0.071 &    0.208 & 0.340   \\
saccade latency x 75~ms  & -0.093 &    0.227 & -0.409\\
saccade latency x 100~ms  &  0.153 &    0.226 & 0.678 \\
\hline
\hline
Random effects variance: Subjects &   \multicolumn{1}{l}{  }  &  \multicolumn{1}{.}{0.2308}  &\multicolumn{1}{l}{}  \\
Random effects variance: Images &   \multicolumn{1}{l}{  }  &  \multicolumn{1}{.}{0.1359}  &\multicolumn{1}{l}{} \\
Log-Likelihood &   \multicolumn{1}{l}{  }  &  \multicolumn{1}{.}{-3990.68}  &\multicolumn{1}{l}{}  \\
Deviance&     \multicolumn{1}{l}{  }  &  \multicolumn{1}{.}{7981.35}  &\multicolumn{1}{l}{}  \\
AIC &     \multicolumn{1}{l}{  }  &  \multicolumn{1}{.}{8011.35}  &\multicolumn{1}{l}{}  \\
BIC &     \multicolumn{1}{l}{  }  &  \multicolumn{1}{.}{8099.75}  &\multicolumn{1}{l}{} \\
N &     \multicolumn{1}{l}{  }  &  \multicolumn{1}{.}{2679}  &\multicolumn{1}{l}{} \\[.2cm]
\multicolumn{4}{r}{ $~^{*} p<.05,~^{**} p<.01,~^{***} p<.001$}

\end{tabular}

\end{table}

%
\subsection{Discussion}
Experiment 3 was conducted to investigate the minimum pre-trial fixation time necessary for a reduction of the early central fixation bias. All pre-trial conditions showed a similar behavior with a tendency of an early CFB as measured by the DTC. We observed the weakest DTC effect for pre-trial fixation times of 125~ms (inset in Fig.~\ref{EXP3}a). Pre-trial fixation times equal to or smaller than 50~ms generated fixation positions closest to the image center. Differences in DTC could be explained by the influence of the first saccade latency on the selection of the second fixation location (Fig.~\ref{EXP3}b). Thus, saccade latencies are the most important factor modulating the CFB. A post-hoc analysis revealed that saccade latencies were only affected in conditions with pre-trial fixation times larger than 50~ms. We conclude that a minimum pre-trial fixation time of around 75~ms is needed to prolong saccade latencies in order to reduce the CFB in scene viewing.

\section{Experiment 4}
In Experiment 4, participants started exploration at the center of the screen. This starting position was chosen to quantify the influence of pre-trial fixation times in a standard scene viewing paradigm.

\subsection{Methods}

\subsubsection{Participants}
In this experiment we recorded eye movements from 10 participants (3 male; 18--36 years old)
 
\subsubsection{Procedure}
Experiment 4 followed the same procedure as the preceding experiments but participants started observation in the center of the screen. We tested pre-trial fixation times of 0, 125 and 250~ms since we observed only subtle changes of results for longer pre-trial fixation times in Experiment 1 and 2. As in Experiment 3, we used a within subject design for the three different pre-trial fixation times such that participants viewed blocks of 40 images for each pre-trial fixation time.

\subsection{Results}

\subsubsection{Distance to Center over Time}

\begin{figure}
\unitlength1mm
\begin{picture}(150,80)
\put(0,0){\includegraphics[width=7cm]{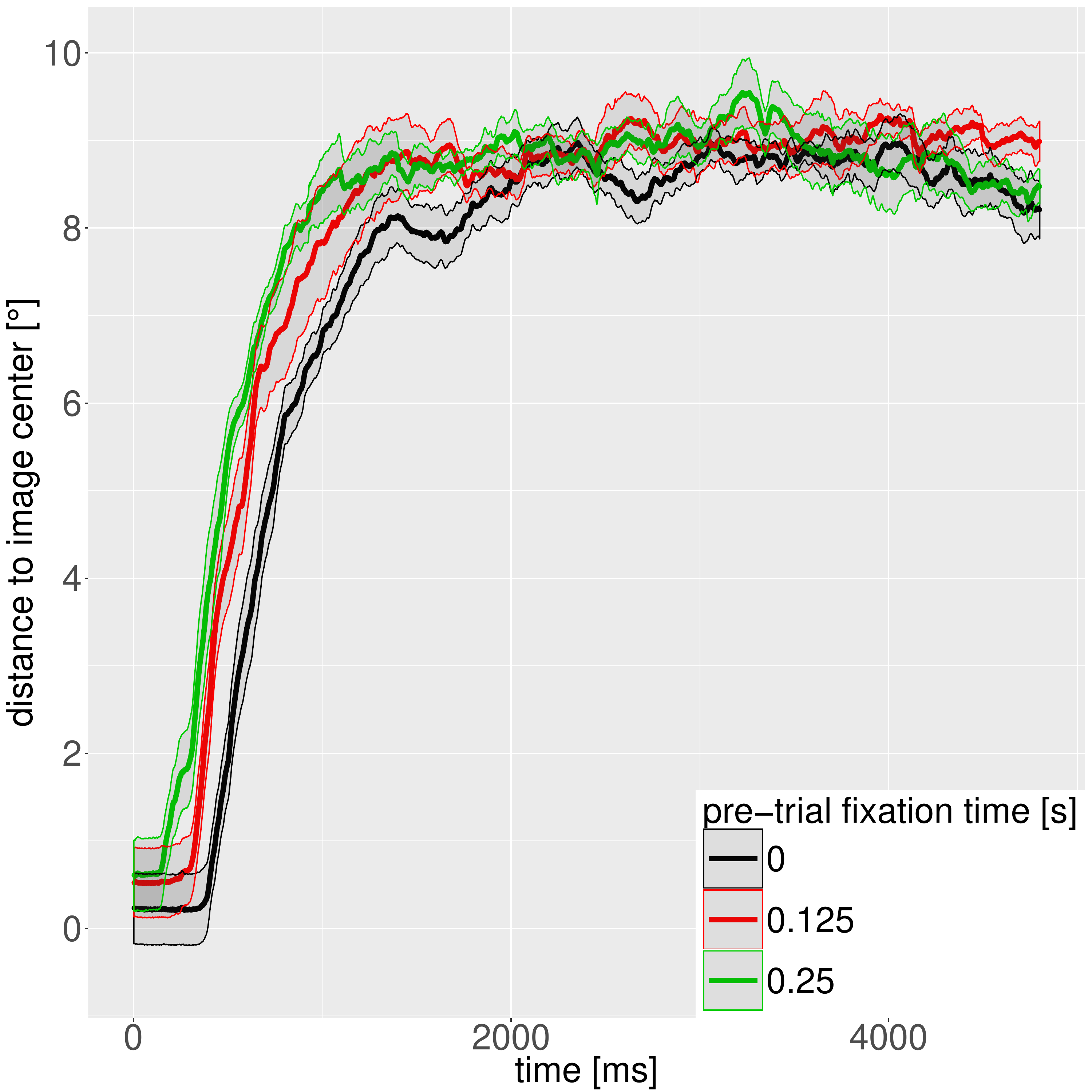}}
\put(80,0){\includegraphics[width=7cm]{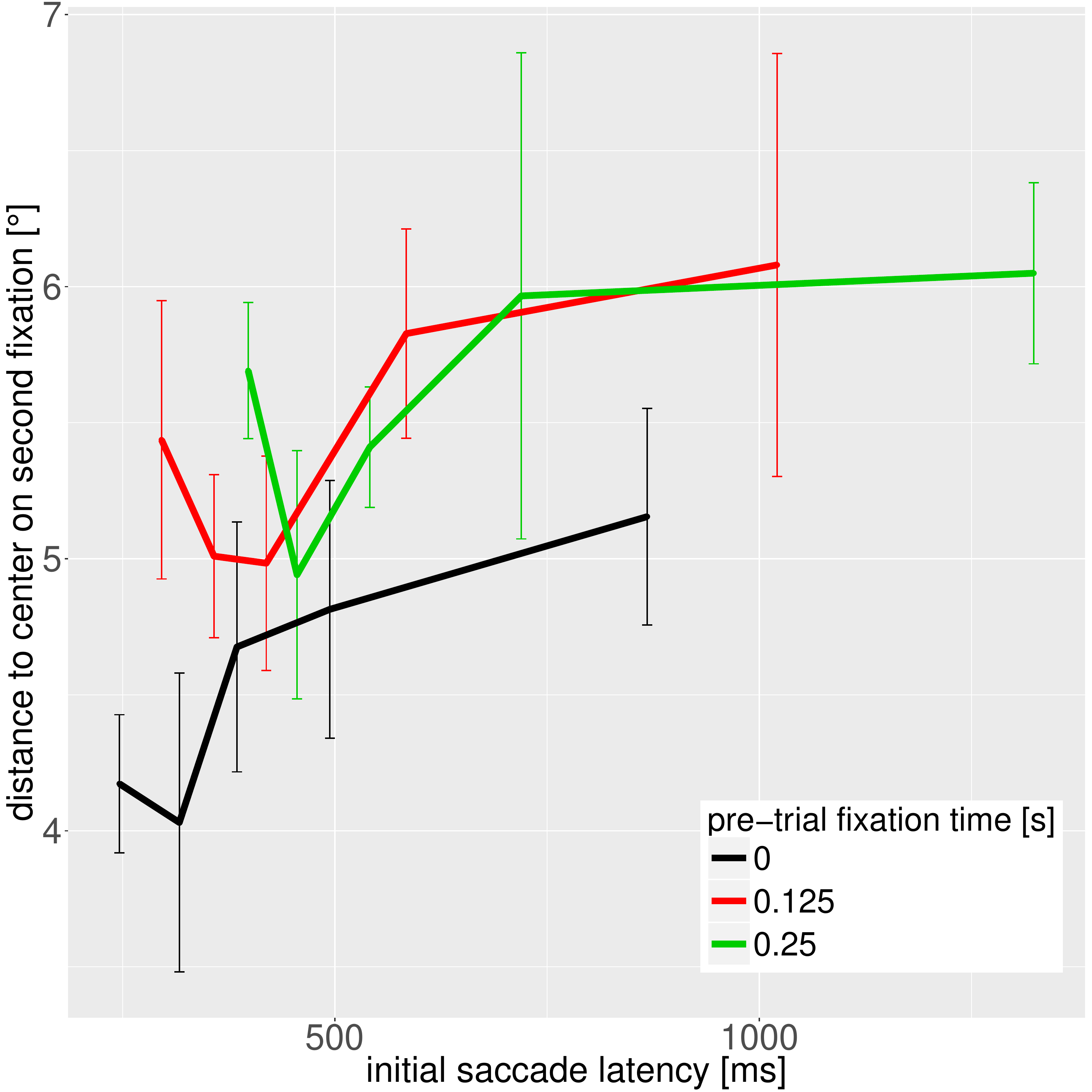}}
\put(0,75)a
\put(80,75)b
\end{picture}
\caption{Experiment 4. a) Mean distance to center over time (${DTC}(t)$) for the three different pre-trial fixation times with starting positions in the center of the image. Confidence intervals indicate standard errors as described by \protect \citeA{cousineau2005confidence}. b) Mean distance to center of the second fixation as a result of initial saccade latency and pre-trial fixation time. Bins represent quintiles of the saccade latency distribution. Errorbars are the standard error of the mean.}
\label{EXP4}
\end{figure}

Contrary to the first experiments initial gaze positions could only move away from the image center with central starting positions in Experiment 4 (Fig.~\ref{EXP4}a).  Therefore, DTC gradually increased until it reached an asymptote. Between pre-trial conditions, DTC differed with respect to the point in time, when curves started to monotonically increase (pre-trial fixation times: 250~ms $<$ 125~ms $<$ 0~ms). Although pre-trial fixation times were chosen to be equidistant, curves for 125~ms and 0~ms pre-trial conditions (red \& black curve) converged later than curves for the 250~ms and 125~ms pre-trial conditions (green \& red curve). This demonstrated that pre-trial fixation times of 125~ms or more reduces the CFB of early fixations even during scene viewing with central starting positions.  

\subsubsection{Distance to Center of the second fixation}

Latencies of the first saccade were longer in this experiment than in any of the other experiments. This observation is in line with results from face perception, where the initial fixation is longer when participants start exploring a face in the center \cite{arizpe2012start}. Due to the increased number of long initial saccade latencies, an influence of saccade latency on the second fixation location was not as clearly visible as in the previous experiments (Fig.~\ref{EXP4}b).
 
Results of a linear mixed model for Experiment 4 partially replicated the main results from Experiment~1--3. The DTC of the second fixation was significantly smaller for a pre-trial fixaiton time of 0~ms.  Ths influence of saccade latency on distance to center of the second fixation did not reach a level of significance of 95\% in Experiment 4. The direction of the influence did stay positive though and nearly reached the level of significance. The fact that saccade latency was not a significant predictor is a result of the rather long latencies and a small number of participants. By removing initial saccade latencies of higher than 1~s (which normally are very rare) saccade latency becomes a significant predictor ($p<0.03$). The interaction between saccade latency and pre-trial times showed that the influence of saccade latency on DTC was, as observed in Experiments 1 and 2, significantly larger for a pre-trial fixation time of 0~ms.

\begin{table}
\caption{Output of LMM for Experiment 4}
\label{LMMExp4}

\small
\begin{tabular}{ @{} R{6.5 cm} . . .}

Fixed Effect   & \multicolumn{1}{c}{ Estimate }  &  \multicolumn{1}{c}{SE}  &\multicolumn{1}{c}{t} \\

\hline 
(Intercept) &     2.858 &   0.157 &  18.194^{***} \\
0~ms 				 &  -0.309 &    0.093 &  -3.312^{***}  \\
125~ms   		&   0.155&   0.083 & 1.877 \\
saccade latency &    0.224 &    0.122 &  1.837 \\
saccade latency x 0~ms  & 0.454 & 0.174 & 2.611^{**} \\
saccade latency x 125~ms &   -0.189 &  0.154 & -1.222 \\
\hline
\hline
Random effects variance: Subjects &   \multicolumn{1}{l}{  }  &  \multicolumn{1}{.}{0.2607}  &\multicolumn{1}{l}{} \\
Random effects variance: Images &   \multicolumn{1}{l}{  }  &  \multicolumn{1}{.}{0.1845}  &\multicolumn{1}{l}{} \\
Log-Likelihood &   \multicolumn{1}{l}{  }  &  \multicolumn{1}{.}{-1877.00}  &\multicolumn{1}{l}{}  \\
Deviance&     \multicolumn{1}{l}{  }  &  \multicolumn{1}{.}{3754.01}  &\multicolumn{1}{l}{} \\
AIC &     \multicolumn{1}{l}{  }  &  \multicolumn{1}{.}{3772.01}  &\multicolumn{1}{l}{}  \\
BIC &     \multicolumn{1}{l}{  }  &  \multicolumn{1}{.}{3817.26}  &\multicolumn{1}{l}{}  \\
N &     \multicolumn{1}{l}{  }  &  \multicolumn{1}{.}{1128}  &\multicolumn{1}{l}{}  \\[.2cm]
\multicolumn{4}{r}{ $~^{*} p<.05,~^{**} p<.01,~^{***} p<.001$}

\end{tabular}

\end{table}


\subsection{Discussion}
In our last experiment we investigated the effect of pre-trial fixation times on the CFB in a standard scene viewing experiment where participants start exploration from the image center. As expected DTC increased in all conditions continuously until it reached an asymptote. The point in time when DTC started to increase varied for different pre-trial fixation times.  We measured the earliest response for pre-trial fixation times of 250~ms and the slowest response after no pre-trial fixation times (0~ms). If we remove latencies of higher than 1 ~s we  can replicate an influence of saccade latencies on DTC of the second fixation. In general saccade latency seems to be a strong mediating factor of the CFB. In addition, we observed long initial saccade latencies when participants started at the image center. This is particularly worrying, because the first fixation is usually omitted from analyses in scene viewing experiments. 

When comparing Experiment 4 to the remaining experiments, the CFB was strongest when participants started at the image center without pre-trial fixation time (0~ms). Only after about 1~s DTC (\& CFB) was comparable between experiments and pre-trial conditions. Since most scene viewing experiments last five seconds or less \cite<c.f., data sets in MIT saliency benchmark; >{mit-saliency-benchmark} a substantial proportion of fixations is biased towards the center during a standard scene viewing experiment. A combination of a non-zero pre-trial fixation time and adjustments of the starting position will reduce the CFB and may help to better understand target selection during scene viewing. We will further comment on this issue in the general discussion.

\begin{figure}[ht]
\center
\includegraphics[width=7cm]{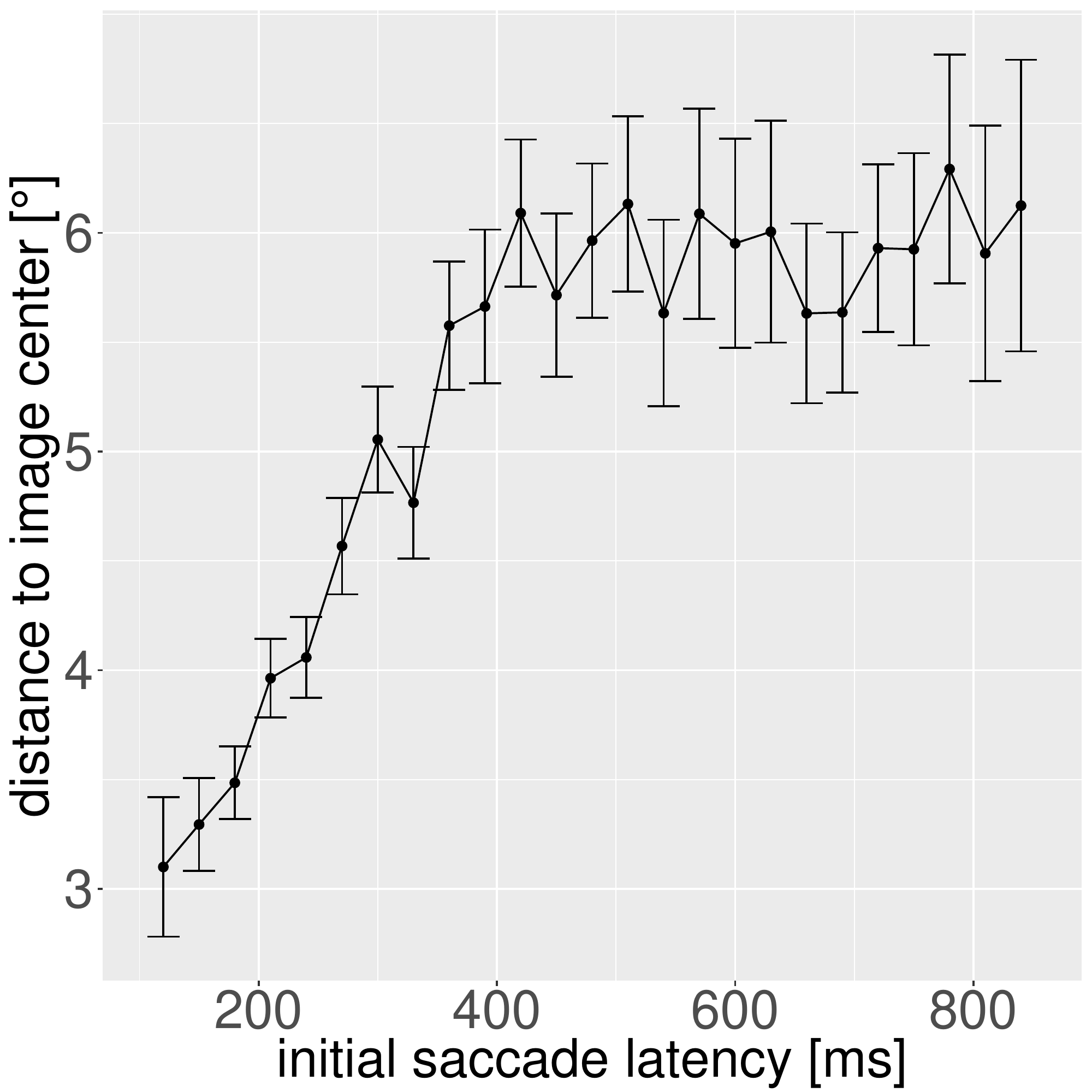}
\label{Fig10}
\caption{Influence of initial saccade latency on the distance to image center of the second fixation for all 4 experiments combined.}
\end{figure}

Figure \ref{Fig10} shows the influence of initial saccade latency on distance to image center for all 4 experiments combined. A clear increase of DTC is visible between 150 and 400~ms. Because initial saccade latencies above 400~ms do not show an influence, pre-trial fixation times above 250~ms did not produce  effects noteworthy. This also explains why in Experiment 4 the rather long saccade latencies were not a significant predictor for the CFB. We conclude our experiments by stating that the initial saccade latency is the dominant factor influencing the early central fixation bias in scene viewing and that by delaying the initial saccade we ensured that the CFB was not exhibited as strongly as in the standard scene viewing paradigm.

\section{Model simulations}

\subsection{Method}

%


Results from our experiments suggest that the early CFB evolves systematically over time with initial saccade latency as a major determinant. To further investigate the time course of the early CFB, we simulated scanpaths generated by different computational models for all experiments and compared simulated scanpaths to our empirical data. Additionally we computed the likelihood of the empirical data under each model \cite{schutt2016likelihood}. Overall we compared five models. First, we simulated three control models: (i) a model that selects fixations proportional to the densities of the empirically observed fixations, (ii) a model which multiplies the fixation density with a Gaussian around the current fixation before selecting the next fixation, (iii) a model that selects fixations from a central activation map to produce a pure CFB. In a second step, we simulated two versions of the dynamical SceneWalk model \cite{engbert2015spatial}: (iv) the original model and (v) a version of the original model with a modulation of the activation map by a sudden image onset. 

For the sampling from a central activation map model (iii) and the extended SceneWalk model  (v) parameters had to be estimated. We used a standard optimization algorithm (fminsearch) implemented in MATLAB \cite{MATLAB:2015} to obtain the parameters with maximum likelihood \cite{bickel2015mathematical,schutt2016likelihood} of fixations 2--4 of half of the participants (Exp.~1--4: $N=20/10/12/5$) and a quarter of the images ($N=30$). We estimated parameters from the second to fourth fixation only for efficiency reasons and since ${DTC}$ curves reached a stable value for later fixations. All models were implemented on a grid of 128$\times$128 cells. For all models we normalized activation maps for target selection to a sum of one to obtain a probability value for each grid cell. For each empirical scanpath, we computed a simulated scanpath with the same number of fixations and fixation durations.

\subsubsection{Density Sampling}
As the most straightforward statistical control, we simulated scanpaths by randomly sampling from a 2D density map of all fixations on a given image, i.e. the empirical fixation map generated by all participants ($N=94$). First, we applied kernel density estimation using the SpatStat package \cite{SpatStat} of the R Language for Statistical Computing \cite{R}. Based on a Gaussian kernel function with a bandwidth parameter of $\lambda=1.05$ (smoothing parameter of the original publication of our SceneWalk model) we computed the empirical fixation density map for each image. Second, to simulate a scanpath (i.e., a fixation sequence), we sampled randomly from this map proportional to the local density so that the normalized density at a particular location translated into the probability to generate a fixation at that position.

\subsubsection{Gaussian Model}
Next, we implemented a statistical model that sequentially sampled from the empirical fixation map via a Gaussian-shaped aperture around fixation to mimic the limits of visual acuity and the attentional span. For a given fixation position $x$, the empirical fixation map was weighted by a two dimensional Gaussian centered around $x$, with a standard deviation of $4.88 \degree$ visual angle. The same standard deviation was used for the size of the attention map of the SceneWalk model by Engbert et al. (see after next section). Fixation locations were sampled from this combined map. To generate a scanpath in this model, the map was recomputed after each fixation. 

\subsubsection{Center Bias Model}
Sampling of scanpaths of the pure CFB model was similar to the sampling of the empirical density map model. Contrary to the empirical density, we estimated an elliptical Gaussian around the image center to account for the CFB. Standard deviations for the horizontal and vertical elongation of the Gaussian were estimated as described in the Methods section. The estimated values for $\sigma_{x}$ for the four experiments were $6.9\degree, 6.6\degree, 7.4\degree$ and $6.9\degree$ and for $\sigma_{y}$  $4.3\degree, 4.4\degree, 4.6\degree$ and $4.4 \degree$ indicating that the center bias was elongated in the horizontal direction. This is in agreement with the finding described by \citeA{clarke2014deriving} for an image independent baseline. The largest values were estimated in Experiment 3, where the size of the image in degree of visual angle was larger than the other experiments (see General Methods).

\subsubsection{Scene Walk Model}

This recently published saccade generation model \cite{engbert2015spatial} proposes that eye movements are driven by two different time-dependent neural activation maps. A fixation map memorizes previous fixations and tags visited fixations locations, making them less probable to be fixated again shortly afterwards. Thus, this map serves as an {\sl inhibition of return} mechanism \cite{itti2001computational,klein2000inhibition}. A second map, the attention map, reflects the attentional allocation on the given scene for a speficic fixation position. To compute the attention map, first an intermediate map is computed by multiplying a two dimensional Gaussian distribution centered at the current fixation position with the empirical saliency map of the image to reflect the reduced processing in the periphery. The influence of attention maps from previous fixations declines over time and thus the previous attention map is increasingly replaced by the map of the new fixation. An equivalent mechanism is used to control the dynamics of inhibition, i.e., the fixation map. The attention and inhibition maps prior to the first fixation are set to uniform distributions.  After computation of the two maps for the current fixation position and duration, they are combined by subtracting the fixation map from the attention map to a target map. After the maps are combined, a target is chosen proportional to the relative activations \cite{luce1959individual} of the target map. Thus, positions where the fixation map is high whereas the attention map is low are rarely fixated and vice versa. Model parameters where chosen as in the published version of the SceneWalk model \cite{engbert2015spatial}. Two  differences to the original publication were that i) we only computed new maps for each new fixation instead of computing new maps every $\Delta t=10~ms$. This was done because we chose fixation durations from the experimental data for our simulations and did thus not need to compute the activation maps for other possible fixation durations (see \citeNP{rothkegel2016influence,schutt2016likelihood}) and ii) negative values of the grid and all values smaller than a constant $\eta$ were transformed by an exponential function such that each value on the grid was larger than zero. This was done because the likelihood of a scanpath containing any impossible fixation would be zero. Thus we could not differentiate between models in terms of likelihood otherwise. After transforming these small or negative values the combined target map was normalized again to obtain a sum of 1.

\subsubsection{SceneWalk StartMap}
\begin{figure}[t]
\unitlength1mm
\begin{picture}(150,100)
\put(0,0){\includegraphics[width=16cm]{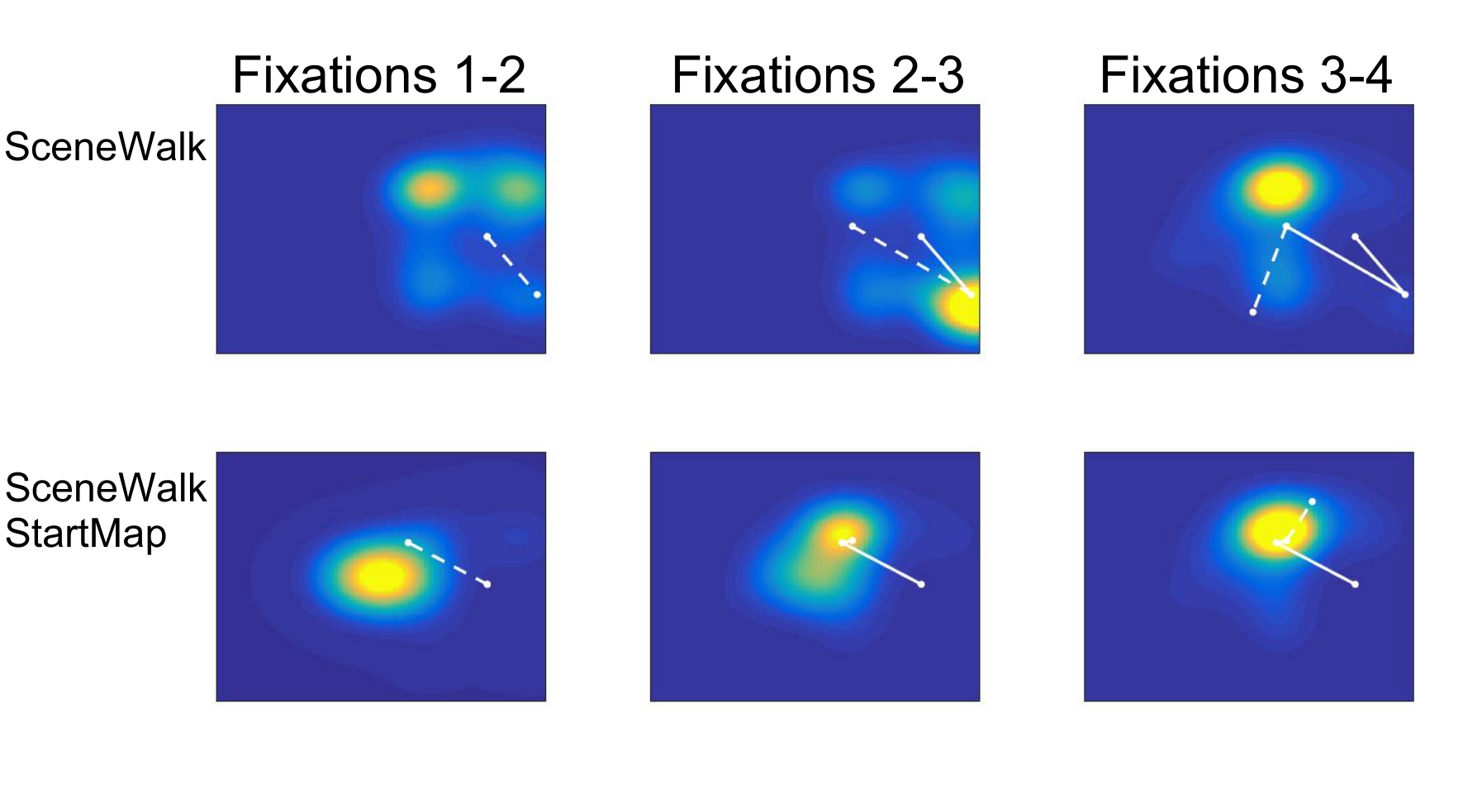}}
\end{picture}
\caption{Fixations 1--4 of a randomly chosen trial with a pre-trial fixation time of 0~ms. In this example the initial fixation duration is very short and thus the initial central activation map of the SceneWalk StartMap has a strong influence on the first saccade target.}
\label{MODMethod}
\end{figure}

Since the original SceneWalk model was not intended to reproduce the early CFB (see results of model simulations), we developed a modified version of the original model that takes the sudden image onset during scene perception into account. We made two changes. First, contrary to the original SceneWalk model with a constant activation across the entire attention map at the beginning of a trial, we decided to use an attention map with higher activations near the center of an image than at the periphery (see Fig.~\ref{MODMethod}b). This was motivated by the sudden image onset that may lead to an initial prioritization of central locations. Second, we realized that the decay of the attention map was too fast during the initial fixation. Therefore, we added a parameter that specified the rate of decay during the initial fixation. For all other fixations we used the same decay parameter as during the original simulations.

The initial center map was an elliptical Gaussian (cf., Center Bias model). We estimated two parameters for the standard deviations of the center map. The horizontal standard deviation $\sigma_{x}$ was estimated at values of $3.5\degree, 1.8\degree$ and $3.9\degree$ for Experiments 1--3. The vertical standard deviation $\sigma_{y}$ for Experiment 1--3 was estimated at $2.3\degree, 2.3\degree$ and $2.4\degree$. The parameters estimated for Experiment 4 were very large with $\sigma_{x} = 136.0\degree$ and $ \sigma_{y} = 4.2\degree$. This resulted in small initial differences in activations between center and periphery for simulations of Experiment 4 and was similar to the constant activations in the original model. The reason for this behaviour arises from the architecture of the model. Since activations in the attention map rise near fixation, central activations are prioritized initially when participants start to explore a scene near the image center. 

%

Figure \ref{MODMethod} shows the simulated fixations 1--4 of a randomly chosen trial for the two different versions of the SceneWalk model. The initial fixation duration in this trial was very short ($t=184~ms$) and thus the initial attention map of the SceneWalk StartMap model is biased stongly towards the center whereas the original SceneWalk model does not show an increased central activation during initial fixations.




\subsection{Results}
We simulated saccadic sequences with the same starting positions, number of fixations, and fixation durations as observed empirically. We computed the ${DTC}$ of saccadic sequences simulated for each experiment (Fig.~\ref{MOD1} \& \ref{MOD2}a), compared the effect of saccade latencies on DTC of the second fixation (Fig.~\ref{MOD2}b), and computed the information gain for Fixation 2--10 for the different models (Fig.~\ref{MOD4}). The initial fixation was excluded from the likelihood estimation since the starting position was determined by the experiment.

\subsubsection{Distance to Center over Time}
\begin{figure}[t]
\unitlength1mm
\begin{picture}(150,100)
\put(0,0){\includegraphics[width=16cm]{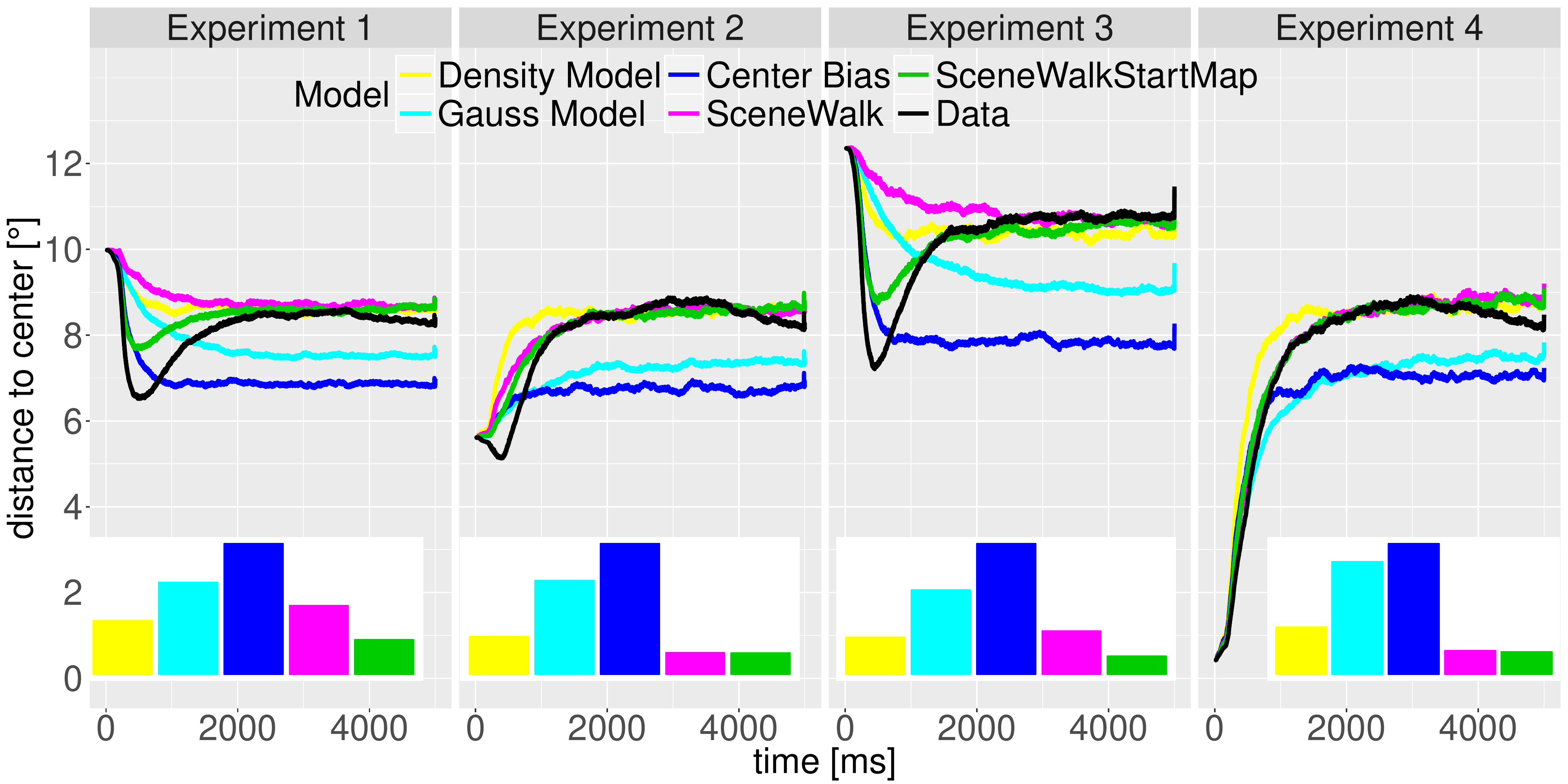}}
\end{picture}
\caption{Distance to image center over time for the empirical data and the 5 saccade generation models. We did not split the data into groups of pre-trial fixation time for visibility purposes. The SceneWalk StartMap model (green curve) is the only one that progresses similarly to the empirical data (black curve). The Gauss model (cyan curve) and the center bias sampling (blue curve) do not converge with the empirical data in the end. Inlays represent the summed deviation of model curves to the empirical data.}
\label{MOD1}
\end{figure}

The average DTC over time is plotted in Figure \ref{MOD1}. We averaged across pre-trial fixation times in a first analysis since DTC curves did not differ between pre-trial fixation times for all models but the SceneWalk StartMap model. In all experiments the SceneWalk StartMap model (green curve) produced a pattern most similar to the empirical data (black curve) but underestimated the empirical early CFB in Experiments~1--3. However, it was the only model to produce the initial characteristic dip of the DTC curves. While the original SceneWalk model (pink curve) was not able to generate an early CFB, it converged with the empirical data after 1-2~s and reached a stable DTC.

The Density model (yellow curve) sampled fixations proportional to the empirical fixation density and was the first model to reach a stable DTC. Since DTC of the Density model and the empirical data converged, we can conclude that fixations across participants were distributed randomly and proportional to the empirical density of all fixation locations on an image after about 2~s. While participants initially explored the center, later fixations were not directed further into the periphery than expected by the overall distribution of fixations. Both the Gaussian model (cyan curve) and Center Bias model (blue curve) moved initially too slowly towards the center and reached a stable DTC after about 1~s that was too close to the center. The bad performance of the Gaussian model demonstrated that a second mechanism is needed to move fixations away from the current fixation location. In the SceneWalk model this is implemented by inhibitory tagging via the fixation map \cite<c.f., >{rothkegel2016influence}. The initial decline of DTC of the Center Bias model was weaker and slower than observed in our experiments (Exp.~1--3). This indicated that the CFB on the second fixation was stronger than estimated from Fixation~2--4. The inlays in Figure \ref{MOD1} show the overall deviation of model curves from the empirical data. The SceneWalk StartMap outperformed other models and deviated the least in all four experiments.

\begin{figure}[t]
\unitlength1mm
\begin{picture}(150,70)
\put(0,0){\includegraphics[width=16cm]{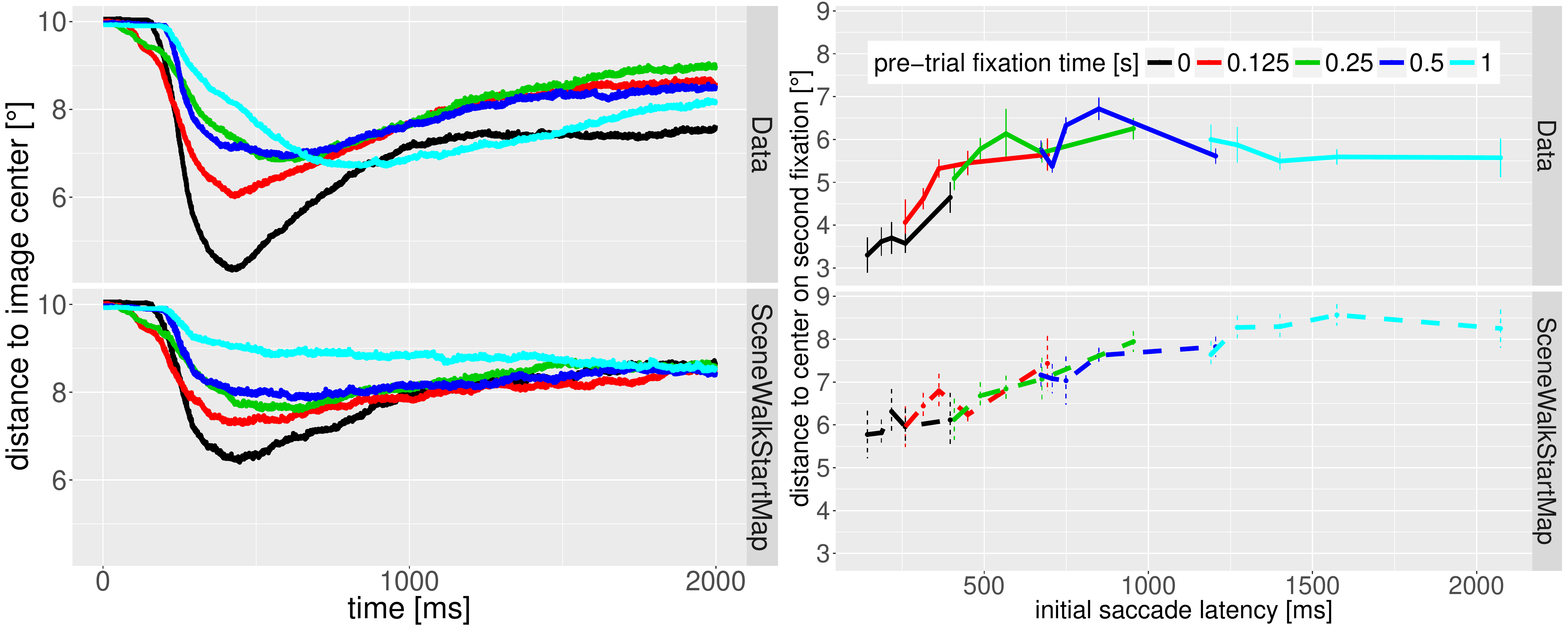}}
\put(0,65)a
\put(80,65)b
\end{picture}
\caption{a) Distance to image center over time for the empirical data and the SceneWalk StartMap model for the different pre-trial fixation times in Experiment 1. b) Influence of the initial saccade latency on the distance to image center on the second fixation for the empirical data and the SceneWalk StartMap model in Experiment 1.}
\label{MOD2}
\end{figure}

Next, we investigated the temporal evolution of the DTC for different pre-trial fixation times for the SceneWalk StartMap model (Fig.~\ref{MOD2}a). The SceneWalk StartMap model only took the initial saccade latency after image onset into account. This produced a qualitatively similar pattern for the different pre-trial fixation times as seen in the data. The qualitative progression for most pre-trial fixation times and experiments was similar to what was observed empirically. It is eminent though that the central fixation tendency produced by the model was too weak when compared to the data. This was probably a result of the method and the fixations used for the parameter estimation (see discussion).

Finally, we evaluated the relation between latencies of the first saccade and DTC of the second fixation (Fig.~\ref{MOD2}b). Again this influence was only visible in the SceneWalk StartMap model as the other models do not incorporate a mechanism that depends on the initial saccade latency. The SceneWalk StartMap model produced a result pattern similar to the empirical data with a similar progresson of lines and a differentiation between pre-trial fixation times. However, the early CFB on the second fixation was too small in all experiments, i.e. the distance to center in all simulations was too large.

\subsubsection{Likelihood}

We computed the likelihood for each model given the empirical data \cite{schutt2016likelihood}. As explained above all model parameters were estimated on Fixation 2 to 4 of half of the participants ($N=5$--$20$) and a quarter of the images ($N=30$) for each experiment. The likelihood was evaluated on the other half of participants and on the rest of the images ($N=90$) for Fixation 2 to 10. All models were computed on the same grid of 128$\times$128 grid cells. The grid cells were normalized to obtain probabilities. This translates to a likelihood for a random fixation on a uniform map of $1/(128\times128) = 2^{-14}$. Thus the information gain for a fixation $i$ at point ($x$,$y$) of a model map $u$ can be computed as
\begin{equation}
\label{Eq_gain}
gain_{i} = \log_2(u[x_i,y_i])+14.
\end{equation}
The computed value represents the information gained relative to a random process that generates a uniform distribution of fixations. 

\begin{figure}
\unitlength1mm
\begin{picture}(150,80)
\put(0,0){\includegraphics[width=16cm]{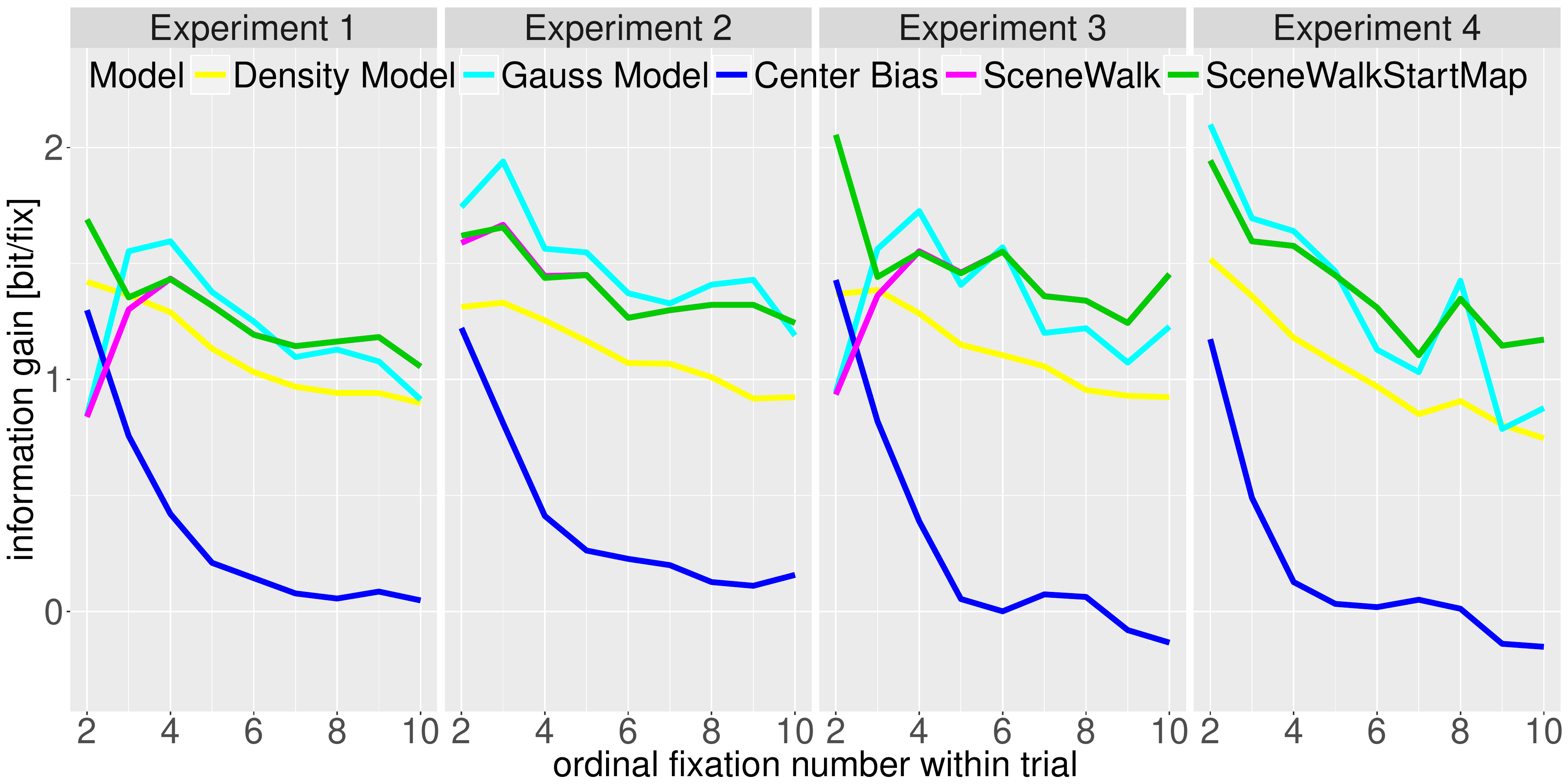}}

\end{picture}
\caption{Mean information gain in bit per fixation for the different models for the Fixation 2 to 10. Fixation 1 is not simulated by the models and set to a likelihood of 1. Higher values of information gain mean that empirically measured fixations are more likely in the corresponding model to occur.}
\label{MOD4}
\end{figure}

Figure \ref{MOD4} shows the information gain of all models for Fixation 2 to 10. The Density model (yellow curve) predicts fixation locations well, but the information gained from this model decreases over time. Thus saliency models could explain less information for later fixations than for the initial ones. The CenterBias explained initial fixations similarly well as the density model, but the gained information declined towards zero for later fixations. Both stationary models were outperformed by all dynamic models on all  fixations except for the second. For this second fixation only the SceneWalk StartMap model (green curve) generated higher information gain than the static models in experiment 1 and 3. The original SceneWalk model (pink curve) performed less good on the second fixation, but makes the same predictions from the third fixation on. Finally, the Gaussian model performed equally well as the SceneWalk models in terms of information gain. Thus we confirm earlier findings that the restriction to nearby saccade targets is the most important influence to include into a model. However, as seen in the previous section, this is not sufficient to explain the DTC over time.

Across experiments and models, the likelihood declined over time. This was due to the larger variability of later fixation locations between participants.

\subsection{Discussion}
Model simulations showed that adjusting the SceneWalk model with an initial map can reasonably explain the central fixation bias. While the information gain of the Gaussian model was similar, the SceneWalk StartMap model was the only model that generated an early dip in DTC curves, qualitatively replicated differences in DTC curves between pre-trial fixation times, and replicated saccade latency effects on DTC of the second fixation.  In particular for starting positions near the border of the screen, a central activation in the attention map was needed to qualitatively reproduce the initial CFB. 

However, the initial CFB generated by the SceneWalk StartMap model was too weak. The same was true for the Center Bias model. This suggests that the procedure to estimate the central bias from the  second to fourth fixation underestimated the initial CFB. Incorporating a stronger bias might further improve predictions of the SceneWalk StartMap model. 

Fixations after about 2~s of exploration were distributed as if drawn from the distribution randomly. Later fixations were not directed towards the periphery more due to the initial CFB. Both SceneWalk models approached a similar DTC during the final viewing period, while the simpler Gaussian model stayed too near to the center.


\section{General Discussion}
During scene viewing the eyes have a strong tendency to fixate near the center of an image which potentially masks other bottom-up and top-down effects of saccadic target selection. In a previous study \cite{rothkegel2016influence} with starting positions near the image border and an experimentally delayed first saccade after the onset of an image we observed a considerable reduction of the central fixation bias \cite<CFB; >{tatler2007central}. Here, we investigated this reduction in four scene viewing experiments. We manipulated starting positions and the latency of the initial saccade. Contrary to the original scene viewing paradigm, where participants started exploration immediately after image onset, we delayed the initial saccadic response by instructing participants to start exploration only after disappearance of a fixation marker. As a measure of the central fixation bias we computed the distance to center (DTC) of the eyes over time. In all experiments the disappearance of a fixation marker 125~ms after image onset led to an early reduction of the DTC in comparison to trials where the fixation marker disappeared simultaneously to the image onset (original scene viewing paradigm). The reduction of the CFB was particularly pronounced in experiments with pre-trial fixation time as a between subject factor (Exp.~1~\& 2). But the reduction of the CFB was even visible when participants started observation at the center of an image (Exp.~4). In a within-subject design with very short pre-trial fixation times (Exp.~3) the manipulation did not significantly reduce the early CFB. A systematic investigation of the disappearance delays of the fixation marker demonstrated that a delay between 125~ms and 250~ms is sufficient for a strong and reliable reduction of the CFB. Shorter delays produced stronger CFBs, whereas longer delays only modestly reduced the CFB but severely changed the time course of the scene viewing paradigm. With long delays the preview time before exploration of an image increased substantially in contrast to the standard scene viewing paradigm.  The distance to center of the second fixation was well predicted by the latency of the initial saccade (time from image onset) across experiments. Short saccade latencies led to a strong bias towards the center whereas longer saccade latencies were less systematically directed towards the center. Hence, the latency of the initial response seemed to primarily account for the observed differences of the CFB. 


Our findings are in agreement with the note communicated earlier that a sudden image onset during scene viewing  represents an artificial laboratory situation and may cause unnatural saccadic behavior \cite<>{marius2009gaze,tatler2011eye}. However, the sudden image onset seems to primarily affect the tendency of the first saccade to move the eyes towards the center of an image. Due to the dependence of fixation locations \cite{engbert2015spatial}, subsequent fixations are more likely located near the center. Hence, only after about 2~s the initial CFB disappeared and fixations were randomly distributed on an image in our experiments. The reason for this reduction, however, remains unknown. Since our manipulation changed the initial saccade latency and reduced the early CFB, the CFB could be a result of an orienting response due to a luminance change or a result of an initial fixation near the center for fast extraction of the gist \cite{tatler2007central}. 





Computational models that aim at predicting the allocation of visual attention on an image are based on the extraction of image features \cite{itti1998model,borji2013state} and top-down cognitive processes \cite{navalpakkam2005attention,cerf2008predicting}. These models are evaluated by comparing human fixations to a weighted distribution of different influences \cite{mit-saliency-benchmark,borji2013state,le2013methods,borji2015salient}. Although bottom-up and top-down influences as well as a combination of the two can predict human fixations \cite{mit-saliency-benchmark}, the CFB is a strong predictor that improves goodness-of-fit more than any other single feature \cite{judd2009learning,mit-saliency-benchmark}. Therefore, most static visual attention models rely heavily on the implementation of a CFB \cite{kummerer2015information}. The early CFB during scene viewing seems to be an automated, stereotyped response of the saccadic system to a sudden image onset. It masks other bottom-up and top-down factors of saccade target selection and its strength critically depends on the duration of a trial since it primarily affects early fixations. Therefore, a reduction of the  CFB during scene viewing, as generated by our paradigm, provides a better understanding of target selection and a more rigorous test of visual attention models than the original scene viewing paradigm. At the minimum, the latency of the first saccade needs to be taken into account, since it strongly influences subsequent viewing behavior. 


By extending a recently published model of saccade generation \cite<SceneWalk model; >{engbert2015spatial,schutt2016likelihood} we were able to account for the empirical data. The model is based on two competing pathways that provide potential saccade targets (attention map) and keep track of recently fixated locations (fixation map). To generate a strong early CFB, we needed to assume that the sudden image onset led to a strong central activation in the attention map. The dynamic model was the only model to qualitatively reproduce the CFB and the relation between saccade latency and the distance to center of the second fixation. However, in its current form the model underestimated both effects.  A control model that randomly selected saccade targets from the distribution of empirically observed fixation locations (Density model) performed worse than the SceneWalk model but demonstrated that fixations are distributed randomly and proportional to the empirical distribution on the image after about 2~s. Note, a very similar target selection mechanism is often assumed in saliency models where targets are sampled randomly from a saliency map. By incorporating systematic eye movement tendencies these models improve \cite<c.f., >[]{le2015saccadic}. A model similar to the SceneWalk model but without a fixation map (Gauss Model) performed well in terms of likelihood but did not reproduce the temporal evolution of the CFB or influences of saccade latencies. Finally, a pure central fixation bias model (CenterBias model) performed poorly on all measures. 

Our results imply to use a modified version of the scene viewing paradigm to study bottom-up and top-down processes of target selection beyond the CFB. To minimize the influence of the sudden image onset, we suggest to use a fixation marker that disappears about 125~ms after image onset. In addition, due to the dependence of successive fixations, scene exploration should not exclusively start near the image center. Instead initial fixations (fixation markers) should be evenly distributed across the entire image or even with a preference towards the periphery. Central parts of the image will be fixated when the eyes move towards the other side of an image. Finally, sudden onsets of stimuli are often used in other laboratory tasks as well (e.g., visual search, face perception). To what extend our results generalize to other domains remains an open question but an early initial CFB might also bias initial fixations in these tasks. 

 
%

\subsection{Conclusion}
Delaying the first saccadic response by 125~ms or more relative to image onset reduced the central fixation bias, which is most pronounced during early fixations. The latency of the first saccade after image onset was the main predictor for the distance to image center of the second fixation in all four experiments relatively independent of the time we enforced. Thus our results suggest to use a modified version of the scene viewing paradigm to better understand target selection beyond the central fixation bias.

\section{Acknowledgements}
This work was supported by Deutsche Forschungsgemeinschaft (grants EN 471/13-1 and WI 2103/4-1 to R. E. and F. A. W., resp.). We thank Benjamin Tatler for providing us with the images of natural scenes from his seminal paper about the central fixation bias. 

\bibliography{Library}
\end{document}